\documentclass[11pt]{article}
\usepackage{amssymb}
\usepackage{amsmath}
\usepackage{latexsym}
\usepackage{graphics}
\usepackage{color}
\catcode`@=11
\renewcommand{\section}
{\@startsection{section}{1}{0pt}{\medskipamount}{\medskipamount}{\large\bf}}
\makeatletter\renewcommand{\subsection}
{\@startsection{subsection}{2}{\z@}{-3.25ex plus -1ex minus -.2ex}
{1.5ex plus .2ex}{\it }}

\numberwithin{equation}{section}
\catcode`@=12

\def\b{\beta}

\def\p{\phi}

\def\sfrac#1#2{{\textstyle\frac{#1}{#2}}}
\def\m{\mu}
\def\n{\nu}
\def\pa{\partial}

\def\beq{\begin{equation}}
\def\eeq{\end{equation}}
\def\bea{\begin{eqnarray}}
\def\eea{\end{eqnarray}}

\newcommand{\im}{\,\mathrm{i}\,}
\newcommand{\diff}{\mathrm{d}}

\newcommand{\R}{{\mathbb{R}}}

\newcommand{\C}{{\mathbb{C}}}
\newcommand{\Z}{{\mathbb{Z}}}

\newcommand{\Idd}{\mathbf{1}}

\newcommand{\Ecal}{{\cal E}}

\newcommand{\Gcal}{{\cal G}}

\newcommand{\yb}{{\bar{y}}}

\newcommand{\ca}{{\cal{A}}}
\newcommand{\cf}{{\cal{F}}}

\newcommand{\man}{{\cal M}}
\newcommand{\cliff}{{{\rm C}\ell}}

\newcommand{\Lcal}{{\cal L}}
\newcommand{\Tr}{{\rm Tr}}

\newcommand{\tr}{{\rm tr}}

\newcommand{\su}{{{\rm SU}(2)}}
\newcommand{\uo}{{{\rm U}(1)}}
\newcommand{\SU}{{{\rm SU}}}

\newcommand{\uk}{{{\rm U}(k)}}

\newcommand{\spin}{{\rm Spin}}

\newcommand{\mphi}{{{\mbf\phi}^{~}_{(m)}}}

\newcommand{\mA}{{\mbf A^{(m)}}}
\newcommand{\ma}{{\mbf a^{(m)}}}
\newcommand{\mF}{{\mbf F^{(m)}}}
\newcommand{\mf}{{\mbf f^{(m)}}}
\newcommand{\mup}{{{\mbf\Sigma}^{~}_{(m)}}}
\newcommand{\mbf}[1]{{\boldsymbol {#1} }}

\def\Dirac{{D\!\!\!\!/\,}} 
\def\Diraccal{{{\cal D}\!\!\!\!/\,}} 

\def\>{\rangle}
\def\<{\langle}
\def\+{\dagger}
\def\={\ =\ }

\begin{document}
\begin{titlepage}
\setcounter{page}{0}
\begin{flushright}
DIAS--09--01\\
HWM--09--01\\
EMPG--09--01\\
\end{flushright}

\vskip 1.8cm

\begin{center}

{\Large\bf Dimensional Reduction, Monopoles \\[10pt] and
  Dynamical Symmetry Breaking}

\vspace{15mm}

{\large Brian P. Dolan${}^{1,2}$} \ \ and \ \ {\large Richard
  J. Szabo${}^3$} 
\\[5mm]
\noindent ${}^1${\em School of Theoretical Physics, Dublin Institute
  of Advanced Studies \\ 10 Burlington Road, Dublin 4, Ireland}
\\[5mm]
\noindent ${}^2${\em Department of Mathematical Physics, National
  University of Ireland\\ Maynooth, Co. Kildare, Ireland}
\\[5mm]
\noindent ${}^3${\em Department of Mathematics and Maxwell Institute
  for Mathematical Sciences\\ Heriot-Watt University,
Colin Maclaurin Building, Riccarton, Edinburgh EH14 4AS, U.K.}
\\[5mm]
{Email: {\tt bdolan@thphys.nuim.ie , R.J.Szabo@ma.hw.ac.uk}}

\vspace{15mm}

\begin{abstract}
\noindent
We consider $\su$-equivariant dimensional reduction of
Yang-Mills-Dirac theory on manifolds of the form $M\times\C P^1$, with
emphasis on the effects of non-trivial magnetic flux on $\C P^1$. The
reduction of Yang-Mills fields gives a chain of coupled
Yang-Mills-Higgs systems on $M$ with a Higgs  potential leading to
dynamical symmetry breaking, as a consequence of the monopole
fields. The reduction of $\su$-symmetric fermions gives massless Dirac
fermions on $M$ transforming under the low-energy gauge group with
Yukawa couplings, again as a result of the internal $\uo$ fluxes. The
tower of massive fermionic Kaluza-Klein states also has Yukawa
interactions and admits a natural $\su$-equivariant truncation by
replacing $\C P^1$ with a fuzzy sphere. In this approach it is
possible to obtain exactly massless chiral fermions in the effective
field theory with Yukawa interactions, without any further
requirements. We work out the spontaneous symmetry breaking patterns
and determine the complete physical particle spectrum in a number of
explicit examples.

\end{abstract}

\end{center}
\end{titlepage}

\section{Introduction}

\noindent
Various schemes have been used to suggest that the Higgs and Yukawa
sectors of the standard model of particle physics may find their
natural origin in a higher-dimensional gauge theory. The natural
candidates for compact internal spaces in such Kaluza-Klein models are
coset spaces $G/H$, as the action of the isometry group $G$ can be
elegantly compensated by gauge transformations in such a way that the
Lie derivative with respect to a Killing vector becomes a gauge
generator. This provides a unification of the gauge and Higgs sectors
in higher dimensions, while the coupling of fermions
to the higher-dimensional gauge theory naturally induces Yukawa
couplings after dimensional reduction. The pioneering scheme realizing
these constructions is called ``coset space dimensional
reduction''~\cite{FMT,KZ1}. It has also been used more
recently for the dimensional reduction of ten-dimensional
supersymmetric gauge theories to four-dimensional field theories with
softly broken $\mathcal{N}=1$ supersymmetry~\cite{Lopes1}, and for the
reduction of superstring theories on nearly K\"ahler
manifolds~\cite{Lopes2}. On the other hand, a generic problem with
Kaluza-Klein reductions has been that they are unable to generate
chiral gauge theories, without some additional
modifications~\cite{KZ1,ShelterIsland}.

In coset space dimensional reduction, one imposes constraints on the
higher-dimensional fields which ensures that they are invariant under
the $G$-action up to gauge transformations. They amount to studying
embeddings of $G$ or of its closed subgroup $H$ in the gauge group of
the higher-dimensional theory. The solutions of the constraints are
then formally identified with the lowest modes of the Kaluza-Klein
towers of the fields, in a field expansion in harmonics on the compact
coset space $G/H$. However, this scheme does not seem to naturally
allow for the incorporation of topologically non-trivial background
fields on $G/H$ which arise from gauging the holonomy subgroup $H$. It
has been shown in~\cite{DN} that, for certain coset spaces, the
inclusion of non-trivial internal fluxes can induce the chiral
fermionic spectrum of quarks and leptons of the standard model.

In this paper we will study the dimensional reduction of gauge
theories in a way which naturally incorporates the topology of gauge
fields on $G/H$. To distinguish our approach from the more standard
coset space techniques, we will refer to it as ``equivariant
dimensional reduction''. The general formalism is developed
in~\cite{ACGP1,LPS1} and has been used to describe vortices as
generalized instantons of higher-dimensional Yang-Mills
theory~\cite{Garcia1}--\cite{LPS3}, as well as to construct explicit
$\su$-equivariant monopole and dyon solutions of pure Yang-Mills
theory in four dimensions~\cite{group4}. Although similar in spirit to
the coset space dimensional reduction scheme, this approach
systematically constructs the unique field configurations on the
higher-dimensional space which are equivariant with respect to the
internal isometry group $G$ and reduces Yang-Mills theory to a quiver
gauge theory. As in coset space dimensional reduction, there is
\emph{a priori} no relation between the gauge group $\Gcal$ of the
higher-dimensional field theory and the groups $G$ or $H$, and the
resulting gauge group of the dimensionally reduced field theory is a
subgroup of $\Gcal$. This is in contrast to the usual Kaluza-Klein
reductions where the isometry group (or the holonomy group) is
identified with the gauge group.

In the following we analyse in detail the simplest case where
$G=\su$ and $H=\uo$, so that the internal space is the projective line
$\C P^1$. In this case the equivariant dimensional reduction of gauge
fields naturally comes with Dirac monopoles. We will emphasize the
effects of the non-trivial monopole background on the physical
particle spectrum obtained from reduction of a Yang-Mills-Dirac
theory. As usual, the mass scale of the dimensionally reduced field
theory is set by the size of the internal space. We will obtain a
Higgs sector of the lower-dimensional gauge
theory with a Higgs potential that leads to dynamical symmetry
breaking, as a direct consequence of the monopole charges. We work out
the complete physical particle content and masses for a variety of
symmetry hierarchies, including one that entails the 
hierarchy $\SU(3)\to\su\times\uo\to\uo$ in which the second step
is dynamical electroweak symmetry breaking.  
An induced Yukawa sector of the reduced
fermionic field theory naturally emerges, again as a direct result of
the internal fluxes. Starting with massless fermions in higher
dimensions, our dimensional reduction induces both massless and
massive fermions. In particular, it naturally allows for the reduction
to massless chiral fermions without the imposition of any extra
structure. In the case of higher spinor harmonic modes, which generate
massive fermions, we show that replacing the coset $\C P^1$ with a
fuzzy sphere gives a natural $\su$-equivariant truncation of the
fermionic Kaluza-Klein tower while maintaining all quantitative
features of the continuous reduction, by using fuzzy spinor fields and 
a universal Dirac operator. Although the classes of models we present
here are far from being phenomenologically viable ones, they provide
a striking illustration of the utility of equivariant dimensional
reduction and how the systematic incorporation of topologically
non-trivial gauge fields of the holonomy group can have dramatic
implications on the physical particle spectrum of the reduced field
theory, including a non-trivial vacuum selection mechanism.

The organisation of this paper is as follows. In \S\ref{eqdimred} we
describe some general aspects of the $\su$-equivariant dimensional
reduction of gauge and fermion fields over $\C P^1$. In
\S\ref{Eqgauge} we derive the corresponding reduction of the pure
massless Yang-Mills-Dirac action functional. In \S\ref{fundrep} and
\S\ref{adrep} we work out large classes of dynamical symmetry breaking
patterns, identifying the entire physical particle spectrum in each
case. In \S\ref{concl} we summarize our findings, discuss some of the
open problems not addressed in our analysis, and comment on the
possibility of obtaining more physically realistic models using
higher-dimensional homogeneous spaces $G/H$ as the internal space.

\bigskip

\section{Equivariant dimensional reduction over $\C
  P^1$\label{eqdimred}}

\noindent
In this section we will describe the dimensional reduction of gauge
and fermion fields over the internal coset space $\C P^1\cong\su/\uo$
which are invariant under the action of the $\su$ isometry group of
$\C P^1$. It is natural to allow for gauge transformations to
accompany the spacetime $\su$ action~\cite{FMT,KZ1}. An elegant and
systematic way to implement such a reduction is via a bundle theoretic
approach. For more details, see~\cite{LPS1,PS1}. 

\subsection{SU(2)-equivariant bundles}

By the inverse relations of induction and restriction~\cite{BGP},
there is a one-to-one correspondence between $\su$-equivariant complex 
vector bundles $\Ecal\to \man:=M\times\C P^1$ and $\uo$-equivariant
complex vector bundles $E\to M$, where $\su$ acts on the space $\man$
via the trivial action on the manifold $M$ and by the standard (left)
transitive action on the projective line $\C P^1\cong\su/\uo$. The
$\uo$ subgroup of $\su$ also acts trivially on $M$. Assume
that the structure group of the principal bundle associated to $\Ecal$
is $\uk$. Imposing the condition of $\su$-equivariance then means that
we should look for representations of the isometry group $\su$ of $\C
P^1$ inside the ${\rm U}(k)$ structure group, i.e. for conjugacy
classes of homomorphisms $\rho:\su\to{\rm U}(k)$. The dimensional
reduction is thus given by $k$-dimensional unitary representations of
$\su$. Up to isomorphism, for each positive integer $r$ there is a
unique irreducible $\su$-module $\underline{V}_{\,r}\cong\C^r$ of
dimension $r$. Therefore, for each positive integer $m$, the module
\beq
\underline{\cal V}\=\bigoplus_{i=0}^m\,
\underline{V}_{\,k_i} \qquad\textrm{with}\quad
\sum_{i=0}^m\,k_i\=k
\label{genrepSU2Uk}\eeq
gives a representation $\rho$ of $\su$ inside ${\rm
  U}(k)$. The original generic ${\rm U}(k)$ gauge symmetry
  then restricts to the centralizer subgroup of the image $\rho(\su)$
  in ${\rm U}(k)$,
\beq
{\rm U}(k)~\longrightarrow~\prod_{i=0}^m\,{\rm U}(k_i) \ ,
\label{gaugebroken}\eeq
which will be the low-energy gauge group of the dimensionally reduced
field theory on $M$. For structure groups $G$ other than $\uk$, the
homomorphisms $\rho:\su\to G$, and hence the analogs of the
restriction patterns (\ref{gaugebroken}), can be deduced from the
Dynkin diagram of $G$.

The Lie algebra of $\su$ is generated by the three Pauli matrices
\beq
\sigma_3\=\begin{pmatrix}1&0\\0&-1\end{pmatrix} \ , \quad
\sigma_+\=\begin{pmatrix}0&1\\0&0\end{pmatrix} \qquad\mbox{and}\qquad
\sigma_-\=\begin{pmatrix}0&0\\1&0\end{pmatrix}
\label{sl2cmatrices}\eeq
with the commutation relations
\beq
\left[\sigma_3\,,\,\sigma_\pm\right]\=\pm\,2\,\sigma_\pm
\qquad\textrm{and}\qquad \left[\sigma_+\,,\,\sigma_-\right]\=\sigma_3
\ . 
\label{sl2cLie}\eeq
The Lie algebra of the $\uo$ subgroup of $\su$ is generated in this
basis by $\sigma_3$. For each $p\in\Z$ there is a unique irreducible
representation $\underline{S}_{\,p}\cong\C$ of $\uo$ given by
$\zeta\cdot v=\zeta^p\,v$ for $\zeta\in S^1$ and
$v\in\underline{S}_{\,p}$. Since the manifold $M$
carries the trivial action of the group $\uo$, any $\uo$-equivariant
bundle $E\to M$ admits a finite Whitney sum decomposition into
isotopical components as~\cite{Segal} $E=\bigoplus_{p}\,E_{(p)}\otimes
\underline{S}_{\,p}$, where the sum runs over the set of eigenvalues
for the $\uo$-action on $E$ and $E_{(p)}\to M$ are bundles with the
trivial $\uo$-action.

The corresponding $\su$-equivariant bundle ${\cal E}\to M\times\C P^1$
is obtained by induction as
\beq
{\cal E}=\su\times_\uo E \ ,
\label{calEind}\eeq
where the $\uo$-action on $\su\times E$ is given by
$h\cdot(g,e)=(g\,h^{-1},h\cdot e)$ for $h\in\uo$, $g\in\su$ and $e\in
E$. The $\sigma_3$-action on $\Ecal$ is described by the isotopical
decomposition of $E$ above. The rest of the $\su$ action, i.e. the
actions of $\sigma_+$ and $\sigma_-=\sigma_+{}^\dag$, follows from the
commutation relations (\ref{sl2cLie}), which shows that the action of
the generator $\sigma_+$ on $E_{(p)}\otimes\underline{S}_{\,p}$
corresponds to bundle morphisms $E_{(p)}\to E_{(p+2)}$, 
along with the trivial $\sigma_+$-actions on the irreducible
$\uo$-modules $\underline{S}_{\,p}$. Introduce the standard Dirac
$p$-monopole line bundle
\beq
\Lcal^{p}:=\su\times_\uo\,\underline{S}_{\,p}
\label{monbundles}\eeq
over the homogeneous space $\C P^1$, with $\Lcal^p=\Lcal^{\otimes p}$
for $p\geq0$ and $\Lcal^p=(\Lcal^\vee\,)^{\otimes(-p)}$ for
$p<0$ where $\Lcal=\Lcal^1$. Then, for the induced complex vector
bundle (\ref{calEind}) over $M\times\C P^1$ of rank~$k$, the
$\sigma_3$-action is given by the $\uo$-equivariant decomposition
\beq
{\cal E}\=\bigoplus_{i=0}^{m}\,{\cal E}_{i}
\qquad\textrm{with}\quad
{\cal E}_{i}\=E_{i}\boxtimes\Lcal^{p_i} \quad \mbox{and} \quad
p_i\=m-2i \ ,
\label{calEansatz}\eeq
where $E_{i}\to M$ are complex vector bundles of rank $k_i$ with
typical fibre the module $\underline{V}_{\,k_{i}}$ in
(\ref{genrepSU2Uk}), and ${\cal E}_{i}\to M\times\C P^1$ is the
bundle with fibres 
\beq
\bigl({\cal  E}_{i}\bigr)_{(x,\xi)}=
\bigl(E_{{i}}\bigr)_{x}\otimes\bigl(\Lcal^{p_i}
\bigr)_{\xi}
\eeq
for $x\in M$ and $\xi\in\C P^1$. On the other hand, the
$\sigma_+$-action is determined by a {\it chain} 
\beq
0~\longrightarrow~\Ecal_{m}~\xrightarrow{\Phi_m}~
\Ecal_{{m-1}}~\xrightarrow{\Phi_{m-1}}~\cdots~
\xrightarrow{\Phi_2}~\Ecal_{1}~
\xrightarrow{\Phi_1}~\Ecal_{0}~\longrightarrow~0
\label{holchain}\eeq
of bundle morphisms between consecutive $\Ecal_{i}$'s. After fixing
hermitean metrics on the complex vector bundles $\Ecal_i\to \man$, the
$\sigma_-$-action is described by reversing the arrows in
(\ref{holchain}) and using the adjoint bundle morphisms
$\Phi_i{}^\dag$.

This decomposition can be understood as follows. Given any
finite-dimensional representation $\underline{V}$ of $\uo$, the
corresponding induced, homogeneous hermitean vector bundle over
the coset space $\C P^1\cong\su/\uo$ is given by the fibred product
\beq
\mathcal{V}=\su\times_\uo\,\underline{V} \ .
\label{indhermbungen}\eeq
Every $\su$-equivariant bundle of finite rank over $\C P^1$, with
respect to the standard transitive action of $\su$ on the homogeneous
space, is of the form (\ref{indhermbungen}). If $\underline{V}$ is
irreducible, then $\uo$ is the structure group of the associated
principal bundle. We consider those representations $\underline{V}$
which descend from some irreducible representation of $\su$ by
restriction to the $\uo$ subgroup. Then the bundle decomposition
(\ref{calEansatz}) is associated with the restriction of the
irreducible $\su$-representation of dimension $r=m+1$.

\subsection{Invariant gauge fields}

Let $M$ be a manifold of real dimension $d$ with local
real coordinates $x=(x^\m )\in\R^{d}$, where the indices $\m ,\n
,\ldots$ run through $1,\ldots ,d$. The projective line $\C P^1$ is a
complex manifold with local complex coordinate $y\in\C$ and its
conjugate $\yb$. The metric
\beq
\diff s^2=\Gcal_{AB}~\diff x^{A}\otimes\diff x^{B}
\label{metrichat}\eeq
on $\man=M\times \C P^1$ will be taken to be the direct product of a chosen
riemannian metric on $M$ and the standard ${\rm SO}(3)$-symmetric metric on
$\C P^1\cong S^2$, where the indices $A,B,\dots$ run over
$1,\dots,d+2$. In the coordinates above it takes the form
\beq\label{metric3}
\diff s^2 =G_{\mu\nu}\ \diff x^\mu\otimes\diff x^\nu +
\frac{4R^2}{\left(1+y\,\yb\right)^2}\ \diff y\otimes\diff \yb \ ,
\eeq
where $R$ is the radius of the sphere $S^2$. We use conventions in
which the coordinates $x$ and $y$ are dimensionless, while the line
element (\ref{metric3}) has mass dimension~$-2$. More generally, one
may consider warped compactifications of $\man$ with the same
topology, but this doesn't seem to add any new qualitative features to
our ensuing results.

Let $\ca$ be a connection on the hermitean vector bundle $\Ecal\to
M\times\C P^1$ having the form given by
$\ca=\ca_A~\diff x^A$ in local coordinates
$(x^A)$ and taking values in the Lie algebra ${\rm u}(k)$. We
will now describe the $\su$-equivariant reduction of $\ca$ on
$M\times\C P^1$. The spherical dependences are in this case completely
determined by the rank $k$ of the bundle $\Ecal$ and the unique (up to
gauge transformations) $\su$-invariant connections $a_{p}$ on the
monopole line bundles (\ref{monbundles}) having, in local complex
coordinates on $\C P^1$, the forms
\beq\label{f1}
a_{p}= \frac{p}{2\left(1 +y\,\yb
\right)}\, \left(\yb~
\diff y -y~\diff\yb\right) \ .
\eeq
The curvatures of these connections are
\beq
f_{p}\=\diff a_{p}\= - \frac{p}
{\left(1 +y\,\yb\right)^2}~
\diff y\wedge \diff \yb \ ,
\label{f2}
\eeq
and their topological charges are given by the degrees of the
complex line bundles $\Lcal^{p}\to\C P^1$ as
\beq
{\rm deg}~\Lcal^{p}\= \frac{\im}{2\pi}\,
\int_{\C P^1}\,f_{p} \= p \ .
\label{f3}
\eeq
Related to the monopole fields are the unique, covariantly constant
$\su$-invariant forms of types $(1,0)$ and $(0,1)$ on $\C P^1$
given respectively by
\begin{equation}\label{f8}
\beta\= \frac{2~\diff y}{1 +y\,\yb} \qquad \mbox{and}\qquad
\bar{\b}\= \frac{2~\diff \yb}{1 +y\,\yb} \ .
\end{equation}
They respectively form a basis of sections of the canonical line
bundles $K=\Lcal^2$ and $K^{-1}=\Lcal^{-2}$, which are the summands of
the complexified cotangent bundle $T^*\C
P^1\otimes\C=K\oplus K^{-1}$ over $\C P^1$. The
$\su$-invariant K\"ahler $(1,1)$-form on $\C P^1$ is $\frac\im2\,
R^2\,\beta\wedge\bar\beta$.

With respect to the isotopical decomposition (\ref{calEansatz}), the
twisted ${\rm u}(k)$-valued gauge potential $\ca$ splits into
$k_{i}\times k_{j}$ blocks $\ca=\left(\ca^{ij}\right)$ with
$\ca^{ij}\in\mbox{Hom}\bigl(\,\underline{V}^{}_{\,k_{j}}\,,\,
\underline{V}^{}_{\,k_{i}}\bigr)$, which we write as
\beq
\ca=\mA(x)\otimes1+\Idd_k\otimes \ma(y)+\mphi(x)\otimes\bar\beta(y)-
\big(\mphi(x)\big)^\dag\otimes\beta(y)
\label{f4}\eeq
where
\beq
\mphi:=\begin{pmatrix}0&\phi_1&0&\dots&0\\0&0&\phi_2&\dots&0\\
\vdots&\vdots&\ddots&\ddots&\vdots\\0&0&0&\dots&\phi_m\\
0&0&0&\dots&0\end{pmatrix}
\label{mgradedphidef}\eeq
while
\beq
\mA~:=~\sum_{i=0}^m\,A^i\otimes\Pi_i \qquad \mbox{and} \qquad
\ma~:=~\sum_{i=0}^m\,a_{p_i}\otimes\Pi_i
\label{mAdef}\eeq
with $\Pi_i:{\cal E}\to{\cal E}_i$ the canonical orthogonal
projections of rank one onto the sub-bundles ${\cal E}_i$, obeying
$\Pi_i\,\Pi_j=\delta_{ij}~\Pi_i$. The bundle morphisms
$\Phi_{i+1}:=\ca^{i\,i+1} = \phi_{i+1}(x)\otimes\bar{\b}(y)\in{\rm
  Hom}(\Ecal_{i+1},\Ecal_{i})$ obey ${\Phi_{m+1}}=0=\Phi_{0}$. The
gauge potentials $A^{i}\in{\rm u}(k_{i})$ are connections on the
hermitean vector bundles $E_{{i}}\to M$. The bifundamental
scalar fields $\phi_{i+1}\in{\rm Hom}(E_{i+1},E_i)$ can be identified
with sections of the bundles $E_i\otimes E_{i+1}^\vee$ and transform
in the representations
$\underline{V}_{\,k_{i}}\otimes\underline{V}_{\,k_{i+1}}^\vee$ of the
subgroups ${\rm U}(k_{i})\times{\rm U}(k_{i+1})$ of the original $\uk$
gauge group. The gauge potential $\ca$ given by
(\ref{f4}) is anti-hermitean and ${\rm SO}(3)$-invariant. All fields
$(A^i, \p_{i+1})$ are dimensionless and depend only on the coordinates
$x\in M$. Every $\su$-invariant unitary connection $\ca$ on $M\times\C 
P^1$ is of the form given in (\ref{f4}) (up to gauge
transformations)~\cite{PS1,BGP}.

The curvature two-form $\cf =\diff\ca + \ca\wedge\ca$ of the
connection $\ca$ has components which are given by
$\cf_{AB}= \pa_A\ca_B-
\pa_B\ca_A+ [\ca_A,\ca_B]$ in local
coordinates $(x^A)$, where $\pa_A:=\pa /\pa
x^A$. It also takes values in the Lie algebra ${\rm u}(k)$,
and in local coordinates on $M\times\C P^1$ it can be written as
\begin{equation}
{\cf}=\sfrac{1}{2}\,{\cf}_{\m\n}~\diff x^{\m}\wedge\diff x^{\n} +
{\cf}_{\m y}~\diff x^{\m}\wedge\diff y +
{\cf}_{\m\yb}~\diff x^{\m}\wedge\diff\yb
+ {\cf}_{y\yb}~\diff y\wedge\diff\yb \ .
\label{curvprod}\end{equation}
The calculation of the curvature (\ref{curvprod}) for $\ca$ of the
form (\ref{f4}) yields
\begin{equation}\label{f10}
\cf\=\left(\cf^{ij}\right) \qquad\mbox{with}\quad
\cf^{ij} \= {\diff}\ca^{ij} +
\sum_{l=0}^{m}\,\ca^{il}\wedge \ca^{lj}
\ ,
\end{equation}
giving
\beq
\cf=\mF+\mf+\big[\mphi\,,\,\mphi^\dag\,\big]~\beta\wedge\bar\beta+
D\mphi\wedge\bar\beta-\big(D\mphi\big)^\dag\wedge\beta
\label{f11}\eeq
where
\beq
\mF~:=~\diff\mA+\mA\wedge\mA \qquad \mbox{and} \qquad 
D\mphi~:=~\diff\mphi+\big[\mA\,,\,\mphi\big] \ ,
\label{mFdef}\eeq
while $\mf=\diff\ma=\sum_i\,f_{p_i}\otimes\Pi_i$ are the
contributions from the monopole fields. We have suppressed the tensor 
product structure pertaining to $M\times \C P^1$. From (\ref{f11}) we
find the non-vanishing field strength components
\bea\label{f15}
\cf^{ii}_{{\mu}{\nu}}&=&F_{{\mu}{\nu}}^i \ , \\[4pt]
\label{f16}
\cf^{i\,i+1}_{{\mu}\yb}&=&\frac{2}{1+y\,{\yb}}\,
D_{{\mu}} \phi_{i+1}~=~-\bigl(\cf^{i+1\,i}_{{\mu}y}\bigr)^\+ \ ,
\\[4pt] \label{Fyyb}
\cf^{ii}_{y\yb}&=& - \frac{1}{(1 +y\,{\yb})^2}\, 
\left(p_i+4\phi_i^\+\,\phi^{~}_i -
  4\phi^{~}_{i+1}\,\phi^\+_{i+1}\right) 
\ ,
\eea
where $F^{i}=\diff A^{i} + A^{i}\wedge
A^{i}=\frac12\,F_{\mu\nu}^{i}~\diff x^\mu\wedge\diff x^\nu$ are the
curvatures of the bundles $E_{{i}}\to M$, and
\begin{equation}\label{der}
D \phi_{i+1}= \diff\phi_{i+1} + A^i\,\phi_{i+1} -
\phi_{i+1}\,A^{i+1}
\end{equation}
are bifundamental covariant derivatives.

The gauge field (\ref{f11}) can be formally identified with the lowest
$\su$-singlet mode in a harmonic expansion of forms on the internal
space $\C P^1$. Since the monopole fields are given by
$f_{p_i}=-\frac{p_i}4\,\beta\wedge\bar\beta$, it can be uniquely
characterized by the requirement that it lives in the kernel of the
covariant derivative operator on $\C P^1$ in the monopole background,
owing to the relations 
\beq
\diff\bar\beta-a_{-2}\wedge\bar\beta\=0\=\diff\beta-a_2\wedge\beta \
. 
\label{covconstrels}\eeq
Equivalently, it is a zero mode of the covariant Laplace operator
acting on forms on $\C P^1$. As usual in Kaluza-Klein reductions,
there is an infinite tower of massive harmonic modes on $M$ which can
also be considered. Their contributions will not be analysed in this
paper.

\subsection{Symmetric spinor fields\label{symspinor}}

Let $M$ be a spin manifold. When $d=\dim_{\R}(M)$ is even, the
generators of the Clifford algebra $\cliff(M\times \C P^1)$ obey 
\beq
\Gamma^A\,\Gamma^B+
\Gamma^B\,\Gamma^A\= 
-2\,\Gcal^{AB}~\Idd_{2^{d/2+1}}
\qquad\mbox{with}\quad A,B\=1,\dots,d+2 \ .
\label{2n2Cliffalg}\eeq
The gamma-matrices in (\ref{2n2Cliffalg}) may be decomposed as
\beq
\bigl\{\Gamma^A\bigr\}\=\bigl\{\Gamma^\mu,\Gamma^y,
\Gamma^\yb\bigr\} \qquad\mbox{with}\quad \Gamma^\mu\=\gamma^\mu
\otimes\Idd_2 \ , ~~ \Gamma^y\=\gamma\otimes\gamma^y
\quad\mbox{and}\quad \Gamma^\yb\=\gamma\otimes\gamma^\yb \ ,
\label{gamma2n2decomp}\eeq
where the $2^{d/2}\times2^{d/2}$ matrices
$\gamma^\mu=-(\gamma^\mu)^\dag$ act locally on the spinor module
$\underline{\Delta}\,(M)$ over the Clifford algebra bundle
$\cliff(M)\to M$, 
\beq
\gamma^\mu\,\gamma^\nu+\gamma^\nu\,\gamma^\mu\=-2\,G^{\mu\nu}~
\Idd_{2^{d/2}}
\qquad\mbox{with}\quad \mu,\nu\=1,\dots,d \ ,
\label{2nCliffalg}\eeq
while
\beq
\gamma\=\frac{\im^{d/2}\,\sqrt{G}}{d!}~
\epsilon_{\mu_1\cdots\mu_{d}}\, 
\gamma^{\mu_1}\cdots\gamma^{\mu_{d}} \qquad\mbox{with}\quad
(\gamma)^2\=\Idd_{2^{d/2}} \quad\mbox{and}\quad
\gamma\,\gamma^\mu\=-\gamma^\mu\,\gamma
\label{chiralityop}\eeq
is the corresponding chirality operator. Here
$\epsilon_{\mu_1\dots\mu_{d}}$ is the Levi-Civita symbol with
$\epsilon_{12\cdots d}=+\,1$. 
The action of the Clifford
algebra $\cliff(\C P^1)$ on the spinor module $\underline{\Delta}\,(\C 
P^1)$ is generated by
\beq
\gamma^y\=-\frac1{2R}\,\left(1+y\,\yb\right)\,\sigma_+
\qquad\mbox{and}\qquad 
\gamma^\yb\=\frac1{2R}\,\left(1+y\,\yb\right)\,\sigma_- \ .
\label{CP1Cliffalg}\eeq
The treatment for $d$ odd is similar.

The $\Ecal$-twisted Dirac operator on $\man=M\times\C P^1$
corresponding to the equivariant gauge potential $\ca$ in (\ref{f4})
is given by
\beq
\Diraccal~:=~\Gamma^A\,{\cal D}_A
{}~=~\gamma^\mu\,D_\mu\otimes\Idd_2+\bigl(\mphi\bigr)\,\gamma\otimes
\gamma^\yb\,\beta_\yb-\bigl(\mphi\bigr)^\dag\,\gamma\otimes\gamma^y\,
\beta_y+\gamma\otimes\Dirac_{\C P^1} \ ,
\label{Diracgradeddef}\eeq
where
\beq
\Dirac_{\C P^1}~:=~\gamma^y\,D_y+\gamma^\yb\,D_\yb~=~
\gamma^y\left(\partial_y+\omega_y+\bigl(\ma\bigr)_y\right)+
\gamma^\yb\left(\partial_\yb+\omega_\yb+\bigl(\ma\bigr)_\yb\right)
\label{DiracS2def}\eeq
and $\omega_y,\omega_\yb$ are the components of the Levi-Civita spin
connection on the tangent bundle of $\C P^1$. The $E$-twisted Dirac
operator $\Dirac:=\gamma^\mu\,D_\mu$ on $M$ is defined as
\beq
\Dirac\=\gamma^\mu\,\Big(\partial_\mu+\theta_\mu+\big(\mA
\big)_\mu\Big) \qquad
\mbox{with} \quad \theta_\mu\=
\mbox{$\frac12$}\,\theta_{\mu\nu\lambda}\,\Sigma^{\nu\lambda} \ ,
\label{DiracM}\eeq
where $\theta=\theta_\mu~\diff x^\mu$ is the spin connection on the
tangent bundle of the manifold $M$ and $\Sigma^{\nu\lambda}$ are the
generators of $\spin(d)$. The operator (\ref{Diracgradeddef}) acts on
spinors $\Psi$ which are ${\rm L}^2$-sections of the bundle
\beq
\Psi\=\begin{pmatrix}\Psi^+\\\Psi^-\end{pmatrix}
{}~\in~\bigoplus_{i=0}^m\,\big(E_{i}\otimes
\underline{\Delta}\,(M)\big)\otimes\begin{pmatrix}\Lcal^{p_i+1}\\
\Lcal^{p_i-1}\end{pmatrix}
\label{spinortotgen}\eeq
over $M\times\C P^1$, where $\Lcal^{p_i+1}\oplus\Lcal^{p_i-1}$
are the twisted spinor bundles of rank two over the sphere~$\C
P^1$ and $\Psi^\pm$ are $2^{d/2+1}$ component spinors satisfying
$(\Idd_{2^{d/2}} \otimes \sigma_3)\Psi^\pm = \pm\,\Psi^\pm$.

The equivariant dimensional reduction of massless Dirac spinors on
$M\times\C P^1$ is defined with respect to symmetric fermions on
$M$. Similarly to the scalar fields $\phi_{i+1}(x)$ in (\ref{f4}),
they act as intertwining operators connecting induced representations
of $\uo$ in the $\uk$ gauge group, and also in the spinor module
$\underline{\Delta}\,(M)$ which admits the isotopical decomposition
\beq
\underline{\Delta}\,(M)\=\bigoplus_{i=0}^m\,\Delta_i\otimes
\underline{S}_{\,p_i} \qquad\mbox{with}\quad \Delta_i\={\rm
  Hom}^{~}_{\uo}\big(\,\underline{S}_{\,p_i}\,,\,
\underline{\Delta}\,(M)\big)
\label{spinmoddecomp}\eeq
obtained by restricting $\underline{\Delta}\,(M)$ to representations
of $\uo\subset\spin(d)\subset\cliff(M)$. The $\Delta_i$'s in
(\ref{spinmoddecomp}) are the corresponding multiplicity spaces, and
using Frobenius reciprocity they may be identified as
\beq
\Delta_{i}={\rm Hom}^{~}_\su\Bigl(\,\underline{\Delta}\,(M)\,,\,
{\rm L}^2\big(\C P^1,\Lcal^{p_i}\big)\Bigr) \ .
\label{multspHomG}\eeq

The isotopical decomposition (\ref{spinmoddecomp}) is now realized
explicitly by using (\ref{multspHomG}) to construct symmetric fermions
on $M$ as $\su$-invariant spinors on $M \times\C P^1$. Analogously to
the invariant gauge fields, they belong to the kernel of the Dirac
operator (\ref{DiracS2def}) on $\C P^1$, and after dimensional
reduction will be massless on $M$. One can write 
\beq
\Dirac_{\C P^1}\=\bigoplus_{i=0}^m\,\Dirac_{p_i}\=
\bigoplus_{i=0}^m\,\begin{pmatrix}0&\Dirac^-_{p_i}\\\Dirac^+_{p_i}&0
\end{pmatrix} \ ,
\label{DiracS2decomp}\eeq
where
\bea
\Dirac^+_{p_i}&=&\frac1{2R}\,\big[\left(1+y\,\yb\right)\,
\partial_\yb-\mbox{$\frac12$}\,(p_i+1)\,y\big] \ , \label{DiracS2p}
\\[4pt] \Dirac^-_{p_i}&=&-\frac1{2R}\,\big[\left(1+y\,\yb\right)\,
\partial_y+\mbox{$\frac12$}\,(p_i-1)\,\yb\big] \ .
\label{DiracS2m}\eea
The operator (\ref{DiracS2decomp}) acts on sections of the bundle
(\ref{spinortotgen}) which we write with respect to this
decomposition as
\beq
\Psi=\bigoplus_{i=0}^m\,\begin{pmatrix}\Psi_{(p_i)}^+\\
\Psi_{(p_i)}^-\end{pmatrix} \ ,
\label{Psidecomp}\eeq
where $\Psi^\pm_{(p_i)}$ are ${\rm L}^2$-sections of
$\Lcal^{p_i\pm1}$ taking values in
$\underline{\Delta}\,(M)\otimes\underline{V}_{\,k_i}$ with
coefficients depending on~$x\in M$.

We need to solve the differential equations
\beq
\Dirac^+_{p_i}\Psi^+_{(p_i)}\=0 \qquad\mbox{and}\qquad
\Dirac^-_{p_i}\Psi^-_{(p_i)}\=0
\label{Diracpkernel}\eeq
for the spinors $\Psi^+_{(p_i)}$ and $\Psi^-_{(p_i)}$ in
$\ker\Dirac^+_{p_i}$ and $\ker\Dirac^-_{p_i}$. By using the forms of
the transition functions for the monopole bundles, one easily sees
that the only solutions of these equations which are regular on both
the northern and southern hemispheres of $S^2$ are of the form
\bea
\Psi_{(p_i)}^+&=&\sum_{\ell=0}^{-p_i-1}\,
\psi_{(p_i)\,\ell}(x)\otimes\chi^+_{(p_i)\,\ell}(y,\yb\,)
\qquad\mbox{and}\qquad 
\Psi_{(p_i)}^-\=0 \quad\mbox{for}\quad p_i~<~0
\label{solschargeneg} \ , \\[4pt]
\Psi_{(p_i)}^-&=&\sum_{\ell=0}^{p_i-1}\,
\widetilde\psi_{(p_i)\,\ell}(x)\otimes\chi^-_{(p_i)\,\ell}(y,\yb\,)
\qquad\mbox{and}\qquad 
\Psi_{(p_i)}^+\=0 \quad\mbox{for}\quad p_i~>~0 \ ,
\label{solschargepos}\eea
with
\beq
\chi^+_{(p_i)\,\ell}(y,\yb\,)\=\frac{y^\ell}
{\left(1+y\,\yb\right)^{-(p_i+1)/2}} \qquad \mbox{and} \qquad
\chi^-_{(p_i)\,\ell}(y,\yb\,)\=\frac{\yb\,^\ell}{\left(1+y\,\yb
\right)^{(p_i-1)/2}} \ . 
\eeq
The components, $\psi_{(p_i)\,\ell}(x)$ and
$\widetilde\psi_{(p_i)\,\ell}(x)$ with  $\ell=0,1,\dots,|p_i|-1$, are
Dirac spinors on $M$ which form the irreducible representation
$\underline{V}_{\,|p_i|}\cong\C^{|p_i|}$ of the group $\su$. This is
of course consistent with the fact that the index of the Dirac
operator $\Dirac_p$ is equal to $-p$.

\subsection{Harmonic spinor fields\label{harmspinor}}

In contrast to the bosonic sector, in the following we will find some
noteworthy features of higher Kaluza-Klein modes in the fermionic
sector, so we shall describe them as well for completeness. They
correspond to eigenspinors with non-zero eigenvalues in the spectrum
of the Dirac operator (\ref{DiracS2decomp}) on $\C P^1$, and are the
only surviving fermions in the absence of the monopole background. The
twisted spinor bundle given by $\Lcal^{p_i+1}\oplus\Lcal^{p_i-1}$
admits an infinite-dimensional vector space of symmetric ${\rm
  L}^2$-sections comprised of spinor harmonics $\Psi_{j,p_i}\in\C^2$,
with $p_i=m-2i$~\cite{CFGB}. They are eigenspinors of the Dirac
operator
$\Dirac_{p_i}$,
$\Dirac_{p_i}\Psi_{j,p_i}=\pm\,\frac1R\,\lambda_{j,p_i}\,\Psi_{j,p_i}$,
with eigenvalues
\beq
\lambda_{j,p_i}=\sqrt{\left(j+\frac{1-p_i} 2\right)\,
\left(j+\frac{1+p_i}2\right)}
\label{lambdalqi}\eeq
each of multiplicity 
\beq
d_{j}=2j+1 \ ,
\label{dlqi}\eeq
where $j$ is integral for odd $p_i$ and half-integral for even $p_i$
with $j \ge \frac{|p_i|+1} 2$. After dimensional reduction, this
produces an infinite Kaluza-Klein tower of massive Dirac spinors on
$M$. We decompose the spinors (\ref{Psidecomp}) in this case as
\bea
\Psi_{(p_i)}^+&=&\sum_{j=\frac{|p_i|+1}{2}}^\infty~
\sum_{\ell=0}^{2j}\,\psi_{(j,p_i)\,\ell}(x)\otimes
\chi_{(j,p_i)\,\ell}^+(y,\yb\,) \ , \nonumber\\[4pt]
\Psi_{(p_i)}^-&=&\sum_{j=\frac{|p_i|+1}{2}}^\infty~
\sum_{\ell=0}^{2j}\,\widetilde\psi_{(j,p_i)\,\ell}(x)
\otimes\chi_{(j,p_i)\,\ell}^-(y,\yb\,) \ ,
\label{Psinon0decomp}\eea
where $\chi_{(j,p_i)\,\ell}^\pm$ are the chiral and antichiral
spinors which are sections of $\Lcal^{p_i\pm1}$, and form an ${\rm
  L}^2$-orthogonal system on $\C P^1$ normalized as
$\big\|\chi_{(j,p_i)\,\ell}^\pm\big\|_{{\rm L}^2}=4\pi\,R^2$ with
\beq
\Dirac_{p_i}^\pm\chi_{(j,p_i)\,\ell}^\pm=\frac1R\,
\lambda_{j,p_i}\,\chi_{(j,p_i)\,\ell}^\mp
\label{Diracchi}\eeq
for each $\ell=0,1,\dots, 2j$. The $|p_i|$ zero modes when
$|p_i|\geq1$ are recovered for $j=\frac12\,(|p_i|-1)$.

In contrast to the zero mode sector, this sector of the dimensionally
reduced field theory contains an infinite number of modes on $M$,
indicated by the infinite range of the angular momenta $j$. In this
case a natural $\su$-invariant way of reducing to a finite number of
fermionic field degrees of freedom is to use a fuzzy sphere $\C
P^1_F$~\cite{Madore}, truncated at some finite level $j=j_{\rm max}$,
as the internal space. Dimensional reduction on the fuzzy sphere was
considered in~\cite{ASMMZ}, although only for the case $m=0$ with no
background monopole fields. Kaluza-Klein compactifications on $\C
P^1_F$ including non-trivial internal magnetic flux are studied
in~\cite{AGSZ}, although in a different context than ours and in
somewhat less generality.

The Dirac equation on the fuzzy sphere, and more generally on fuzzy
$\C P^N$, has been analysed in~\cite{GP} and~\cite{DHMC} respectively.
The spectrum of the (universal) fuzzy Dirac operator including
monopole backgrounds for a given maximal angular momentum $j_{\rm
  max}$ consists again of the eigenvalues (\ref{lambdalqi}) as in the
continuous case, except that now $j\leq j_{\rm max}$. The
corresponding eigenspinors can be constructed as finite-dimensional
matrices. With $L=j_{\rm max}+\frac12$, positive chirality spinors
$\hat\chi^+_{(j,p)\,\ell}$ are complex matrices of dimension given by
$\big(L-\frac{p} 2\bigr)\times\big(L+ 1+\frac{p}2\bigr)$, while
negative chirality spinors $\hat\chi^-_{(j,p'\,)\,\ell}$ are matrices
of dimension $\big(L+1-\frac{p'}2\,\big)\times\big(L+
\frac{p'}2\,\big)$. By truncating at $j_{\rm
  max}=\frac{|p_i|-1}{2}$, all spinor
fields $\psi_{(j,p_i)\,\ell}$ and $\widetilde\psi_{(j,p_i)\,\ell}$
vanish identically, and only the finitely many flavours of the zero-mode
symmetric
spinors $\psi_{(p_i)\,\ell}$ and $\widetilde\psi_{(p_i)\,\ell}$ of
\S\ref{symspinor} survive the dimensional reduction.

\bigskip

\section{Equivariant gauge theory of Kaluza-Klein
  modes\label{Eqgauge}}

\noindent
In this section we will work out the equivariant dimensional
reduction of the pure massless Yang-Mills-Dirac action on $M\times\C
P^1$. We will find that the role of the monopole fields on $\C P^1$ is
to induce a Higgs potential with dynamical symmetry breaking, as well
as couplings to massless spinors with Yukawa interactions from the
zero modes of the Dirac operator $\Dirac_{\C P^1}$. The mass scale of
the broken symmetry phase on $M$ is determined by the size $R$ of the
internal coset space. An induced Yukawa sector of the low-energy
effective field theory then emerges with the standard form of
spontaneous symmetry breaking, containing both massless and massive
fermions together with Yukawa interactions with the physical Higgs
fields. In particular, we will unveil the possibility of obtaining
exactly massless chiral fermions on $M$ with Yukawa interactions,
which can be interpreted as multiplets of left-handed quarks. Our
approach thus avoids the extra requirements necessary for obtaining
chiral fermions in the more conventional coset space dimensional
reduction schemes~\cite{KZ1}.

\subsection{Dimensional reduction of the Yang-Mills action}

For the usual Yang-Mills lagrangian
\begin{equation}\label{lagr}
L^{~}_{\rm YM}=-\frac{1}{4\tilde g^2}\,\sqrt{|\Gcal|}~\tr^{~}_{k\times k}\ 
\cf_{AB}\,\cf^{AB}
\end{equation}
on $\man=M\times\C P^1$, one has
\bea
L^{~}_{\rm YM}&=&-\frac{1}{4\tilde g^2}\,\sqrt{|\Gcal|}~
\tr^{~}_{k\times k} \bigg[\,\cf_{\mu\nu}\,\cf^{\mu\nu}+ 
\frac{(1+y\,\bar y)^2}{2R^2}\,G^{\mu\nu}\,
\left(\cf_{\mu y}\,\cf_{\nu \bar y}+\cf_{\mu\yb}\,\cf_{\nu y}
\right) \nonumber\\ && \qquad\qquad\qquad\qquad\qquad
+\,\frac18\,\Big(\frac{\left(1+y\,\yb\right)^2}{R^2}\,
\cf_{y\yb}\Big)^2\,\bigg] \ .
\label{lagrprod}\eea
The $(d+2)$-dimensional $\uk$ Yang-Mills coupling constant $\tilde g$
has the standard mass dimension $1-\frac d2$ in order to make
(\ref{lagrprod}) dimensionless. The dimensional reduction of the
corresponding Yang-Mills action can be obtained by substituting
(\ref{f15})--(\ref{Fyyb}) into (\ref{lagrprod}) and performing the
integral over $\C P^1$ to arrive at the action
\bea
S^{~}_{\rm YM}&:=&\int_{M\times {\C P^1}}\,
{\diff^{d+2}}x~L_{\rm YM}^{~}\nonumber\\[4pt] &=&
\frac{\pi\,R^2}{\tilde g^2}\, 
\int_{M}\,\diff^{d}x~\sqrt{|G|}\ \sum_{i=0}^m
\,\tr^{~}_{k_i\times k_i}\bigg[\bigl(F_{{\mu}{\nu}}^i\bigr)^\+\,
\bigl(F^{i\,{\mu}{\nu}}\bigr)
+ \frac2{R^2}\,\bigl(D_{{\mu}} \phi_{i+1}\bigr)
\,\bigl(D^{{\mu}} \phi_{i+1}\bigr)^\+\nonumber\\ &&+
\frac2{R^2}\,\bigl(D_{{\mu}}
\phi_{i}\bigr)^\+\, \bigl(D^{{\mu}}\phi_{i}\bigr)
+\frac1{8R^4}\,
\left(p_i+4\phi_i^\+\,\phi^{~}_i - 4\phi^{~}_{i+1}
\,\phi^\+_{i+1}\right)^2\,\bigg] \ .
\label{EF}\eea
This result holds irrespectively of the signature of the chosen metric
on the manifold $M$.

The action (\ref{EF}) defines a non-abelian Higgs model describing $m$ 
interacting complex ``rectangular'' scalar fields coupled to $m+1$
non-abelian gauge fields. From (\ref{der}) it follows that the $\uo$
factor in $\uk\approx\SU(k)\times\uo$ does not enter the bicovariant
derivatives of $\phi_{i+1}$, since an overall $\uo$ factor cancels
between the $A^i$ and the $A^{i+1}$ terms. For the purposes of the
ensuing analysis in this subsection we can therefore focus on 
gauge group $\SU(k)$, though the overall $\uo$ subgroup would in
general couple to fermions, in the fundamental representation of
$\SU(k)$ for example. The decomposition (\ref{gaugebroken}) of the
gauge group arising from the regular embedding of $\su$ is then
modified to
\beq
\SU(k)~\longrightarrow~\uo^m\times\prod_{i=0}^m\,\SU(k_i) \qquad
\mbox{with} \quad \sum_{i=0}^m\,k_i\=k \ .
\label{SUkgaugebroken}\eeq
The gauge coupling in $d$ dimensions should have mass dimension
$2-\frac d2$, so we define $g^2=\tilde g^2/4\pi\,R^2$ as
the $d$-dimensional gauge coupling constant. We then rescale
$\phi_i\to g\,R\,\phi_i$ and $A^i\to g\,A^i$ so that the scalar fields
and the gauge fields have the correct canonical dimensions for a
$d$-dimensional field theory (with dimensionless coordinates).

The action (\ref{EF}) can be succinctly rewritten as a matrix model by
using the operators (\ref{mgradedphidef}), (\ref{mAdef}) and
(\ref{mFdef}) (with the rescalings $\mphi\to g\,R\,\mphi$ and $\mA\to
g\,\mA$), together with
\beq
\mup:=\sum_{i=0}^m\,p_i~\Pi_i
\label{mupdef}\eeq
with respect to the decomposition (\ref{calEansatz}). One then has
\beq
S^{~}_{\rm YM}=\int_M\,\diff^dx~\sqrt{|G|}~
\bigg[\,\tr^{~}_{k\times k}\Big(
\mbox{$\frac14$}\,\big(\mF\big)^\+_{\mu\nu}\,
\big(\mF\big)^{\mu\nu}+\big(D_\mu\mphi\big)^\dag\,\big(D^\mu\mphi\big)
\Big)+V\big(\mphi\big)\bigg] \ ,
\label{YMactionmatrix}\eeq
where the Higgs potential is given by
\beq
V\big(\mphi\big)=\frac{g^2}{2}~\tr^{~}_{k\times k}
\left(\frac1{4g^2\,R^2}\,\mup-
\Big[\mphi\,,\,\mphi^\dag\,\Big]\right)^2 \ .
\label{Higgspot}\eeq
The Higgs potential (\ref{Higgspot}) generically leads to dynamical
symmetry breaking, as a direct consequence of the non-trivial monopole
background on $\C P^1$. Its critical points are described by the
matrix equations 
\beq
\mphi- 2g^2\,R^2\,\left[\big[\mphi,\mphi^\dag\,
\big]\,,\,\mphi\right]=0 \ ,
\label{Higgsextr}\eeq
where we have used $\big[\mup\,,\,\mphi\big]=2\mphi$. When they exist,
solutions of the equation 
\beq
\Big[\mphi\,,\,\mphi^\dag\,\Big]=\frac1{4g^2\,R^2}\,\mup
\label{Higgsvacua}\eeq
give the vacua of the Higgs sector of the field theory.

When $k_0=k_1=\dots=k_m=n$, so that the gauge symmetry restriction is
given by
\beq
\SU(k)~\longrightarrow~\uo^m\times\SU(n)^{m+1} \qquad \mbox{with}
\quad k\=n\,(m+1) \ ,
\label{kieqbreak}\eeq
an explicit solution of (\ref{Higgsvacua}) is given by $\mphi=\mphi^0$,
where
\beq
\phi_i^0=\frac{\zeta_i}{2g\,R}\,\sqrt{i\,(m-i+1)}~\Idd_n
\label{phii0}\eeq
for $i=1,\dots,m$ with $\zeta_i\in S^1$ independent phase
factors. The phases $\zeta_i$ can be removed by a $\uo^m$ gauge
transformation in the unbroken symmetry phase. This solution breaks
the gauge symmetry of the $d$-dimensional field theory on $M$ to
$\SU(n)$. In the broken symmetry phase there are $m\,n^2$ massive
gauge bosons, and $m\,n^2$ physical Higgs fields which can be
represented in terms of $n\times n$ hermitean matrices $h_i$,
$i=1,\dots,m$ with $\phi_i=\phi_i^0+h_i$. The corresponding Higgs
masses, proportional to $\frac1R$, can then be worked out by
substitution into the Higgs potential (\ref{Higgspot}), while the
vector boson masses, also proportional to $\frac1R$, can be worked out
by substitution into the covariant derivative terms of the action
(\ref{YMactionmatrix}). Note that for $n=1$, the gauge symmetry
reduction (\ref{kieqbreak}) is to the maximal torus of $\SU(m+1)$, and
all gauge bosons become massive with no residual gauge symmetry
remaining. In the subsequent sections we will look at some explicit
examples of such mass generation in the field theory defined by
(\ref{YMactionmatrix}).

\subsection{Fermionic action for symmetric spinors}

Using the gauged Dirac operator (\ref{Diracgradeddef}), we may define
a euclidean fermionic action functional on the space of ${\rm
  L}^2$-sections of the bundle (\ref{spinortotgen}) by
\beq
S^{~}_{\rm D}:=\int_{M\times\C P^1}\,\diff^{d+2}x~\sqrt {|\Gcal|}~
\Psi^\dag\,\Diraccal\Psi \ , 
\label{EDdef}\eeq
where $\Psi$ has canonical mass dimension $\frac12\,(d+1)$. In
lorentzian signature the adjoint spinor $\Psi^\dag$ should be replaced
by $\overline\Psi:=\frac 1{\sqrt{-\Gcal^{00}}}\,\Psi^\dag\Gamma^0$.
For definiteness, we shall only
consider the case where the spinor field $\Psi$ transforms under the
fundamental representation of the initial gauge group $\SU(k)$. Other
fermion representations of $\SU(k)$ can be treated similarly.

One has
\beq
\Psi^\dag\,\left(\gamma\,\bigl(\mphi\bigr)\otimes\sigma_-+\gamma\,
\bigl(\mphi\bigr)^\dag\otimes\sigma_+\right)\Psi=\begin{pmatrix}
\left(\Psi^+\right)^\dag &\left(\Psi^-\right)^\dag\end{pmatrix}\, 
\begin{pmatrix}\gamma\,\bigl(\mphi\bigr)^\dag\,\bigl(\Psi^-\bigr) \\
\gamma\,\bigl(\mphi\bigr)\,\bigl(\Psi^+\bigr)\end{pmatrix} \ .
\label{PsiPhiid}\eeq
Substituting (\ref{DiracS2decomp})--(\ref{solschargepos}), we see that 
(\ref{PsiPhiid}) vanishes on symmetric spinors if $m$ is even. On the
other hand, if $m$ is odd there  is a surviving contribution from
(\ref{solschargeneg}) and (\ref{solschargepos}) when $p_i=\pm\,1$. For 
$p_i=-1$ the single positive chirality zero mode is a section of the
trivial line bundle ${\cal L}^0$ over $\C P^1$, as is the single
negative chirality zero mode for $p_i=+1$. They are thus globally
defined functions on $\C P^1$, and hence are simply constants in
(\ref{solschargeneg}) and (\ref{solschargepos}) corresponding to
$\su$-singlets in the trivial representation $\underline{V}_{\,1}$.
For these special cases the expression (\ref{PsiPhiid}) produces
Yukawa couplings to the fields $\Psi^\pm_{(\mp\,1)}$ on $M$.

After integration over $\C P^1$, and the rescalings $\phi_i
\rightarrow g\,R\,\phi_i$,
$\psi_{(p_i)\,\ell}\to(4\pi\,R^2)^{-1/2}\,\psi_{(p_i)\,\ell}$, and
$\widetilde\psi_{(p_i)\,\ell}\to
(4\pi\,R^2)^{-1/2}\,\widetilde\psi_{(p_i)\,\ell}$ to give the scalar
and fermion fields the correct canonical dimensions on $M$, the
contribution from fermion zero modes on $\C P^1$ to the action
functional (\ref{EDdef}) is given by
\bea
S_{{\rm D}}^0&=&\int_{M}\,\diff^dx~\sqrt{|G|}~\Biggl[~
\sum_{i=m_+}^m~\sum_{\ell=0}^{|p_i|-1}~\big(\psi_{(p_i)\,\ell}
\big)^\dag\,\Dirac\bigl(\psi_{(p_i)\,\ell}\bigr)+\sum_{i=0}^{m_-}~
\sum_{\ell=0}^{|p_i|-1}~\big(\widetilde\psi_{(p_i)\,\ell}\big)^\dag\,
\Dirac\bigl(\widetilde\psi_{(p_i)\,\ell}\bigr)\Biggr]\nonumber\\
&&+\,\frac g2\,\int_{M}\,\diff^dx~\sqrt{|G|}~
\biggl[\left(\psi_{(-1)}\right)^\dag \,\phi^\dag_{m_+}\,
\gamma\,\widetilde\psi_{(1)}
+\big(\widetilde\psi_{(1)}\big)^\dag\,\phi_{m_+}\, \gamma\,\psi_{(-1)} 
\biggr] \ ,
\label{EFred}\eea
where $m_-=\big\lfloor\frac{m-1}2\big\rfloor$ and
$m_+=\big\lceil\frac{m+1}2\big\rceil$. The second term in
(\ref{EFred}) is present only when $m$ is odd, in which case
$m_+=\frac{m+1} 2$. The fermion fields $\psi_{(p_i)\,\ell}$ and
$\widetilde\psi_{(p_i)\,\ell}$ for each $\ell=0,1,\dots,|p_i|-1$, with
$\psi_{(-1)}:=\psi_{(-1)\,0}$ and
$\widetilde\psi_{(1)}:=\widetilde\psi_{(1)\,0}$, transform in the
fundamental representation of $\SU(k_i)$. Recall that this sector of
the field theory on $M$ is induced entirely by the non-trivial
monopole background on $\C P^1$.

\subsection{Fermionic action for spinor harmonic modes}

For eigenspinors on $\C P^1$ with non-zero eigenvalues, the term
(\ref{PsiPhiid}) produces an infinite chain of Yukawa couplings
to the Higgs fields $\phi_i$. After integration over $\C P^1$,
and again rescaling
\beq 
\phi_i ~\rightarrow~ g\,R\,\phi_i \ , \quad 
\psi_{(j,p_i)\,\ell}~\rightarrow~\big(4\pi\,R^2\big)^{-1/2}\,
\psi_{(j,p_i)\,\ell} \quad \mbox{and} \quad
\widetilde\psi_{(j,p_i)\,\ell}~\rightarrow~
\big(4\pi\,R^2\big)^{-1/2}\,\widetilde\psi_{(j,p_i)\,\ell}\, ,
\label{PhiPsiRescalings}\eeq
the contributions from non-zero fermion modes on $\C P^1$ with
positive eigenvalues are given by 
$\sum_{i=0}^m\,S^{(p_i)}_{{\rm D}}$ with
\bea
S^{(p_i)}_{{\rm D}}&=&\int_{M}\,\diff^{d}x~\sqrt{|G|}~
\sum_{j=j_{\rm min}}^\infty~
\sum_{\ell=0}^{2j}\,\Big[
\left(\psi_{(j,p_i)\,\ell}\right)^\dag\,\Big(\Dirac+
\frac1R\,\lambda_{j,p_i}\,
\gamma\Big) \psi_{(j,p_i)\,\ell}\nonumber\\ &&
\qquad\qquad\qquad\qquad\qquad \qquad\qquad 
+\,\big(\widetilde\psi_{(j,p_i)\,\ell}\big)^\dag\,
\Big(\Dirac+\frac1R\,\lambda_{j,p_i}\,\gamma\Big)
\widetilde\psi_{(j,p_i)\,\ell} \nonumber\\ && 
 + \,\frac{g}{2}\, \bigl(\psi_{(j,p_i)\,\ell}
\bigr)^\dag\,\phi_i^\dag\,\gamma\,
\widetilde\psi_{(j,p_i+2)\,\ell}+\frac{g}{2}\,
\bigl(\widetilde\psi_{(j,p_i+2)\,\ell}
\bigr)^\dag\,\phi_i\,\gamma\,\psi_{(j,p_i)\,\ell} \Bigr] \ , 
\label{nonnegative}
\eea
where $j_{\rm
  min}=\max\bigl(\frac{|p_i|+1}{2},\frac{|p_{i}+2|+1}{2}\bigr)$. Again
the fermion fields $\psi_{(j,p_i)\,\ell}$ and
$\widetilde\psi_{(j,p_i)\,\ell}$, for each $j\geq j_{\rm min}$ and
$\ell=0,1,\dots,2j$, transform in the fundamental representation of
$\SU(k_i)$.

Consider the truncation of the infinite tower of massive Dirac spinors
on the fuzzy sphere $\C P_F^1$ at 
$j=j_{\rm max}$ as described in \S\ref{harmspinor}. The fuzzy analogue
of the ${\rm L}^2$-norm $\big\|\chi_{(j,p)\,\ell}^\pm\big\|_{{\rm
    L}^2}$ on $\C P^1$ is the matrix trace $\Tr
\big[\big(\hat\chi^\pm_{(j,p)\,\ell}\big)^\dag\,\hat
\chi^\pm_{(j,p)\,\ell}\big]$, but observe that
$\Tr\big[\big(\hat\chi^+_{(j,p)\,\ell}\big)^\dag\,\hat
\chi^-_{(j,p'\,)\,\ell}\,\big]$ is also well-defined if and only if $p'=p+2$
and this
is exactly what is needed for the Yukawa couplings in
(\ref{nonnegative}). Note that when $i=0$, the Yukawa terms in
(\ref{nonnegative}) vanish because there is no fermion field
$\widetilde\psi_{(j,m+2)\,\ell}$. Similarly, the minimum value of the
monopole charge $p_i$ in any spinor $\psi_{(j,p_i)\,\ell}$ is $-m$,
when $i=m$, so the fermion fields $\psi_{(j,-m-2)\,\ell}$ are
never present either. Hence $\psi_{(j,m)\,\ell}$ and
$\widetilde\psi_{(j,-m)\,\ell}$ have no Yukawa couplings, because they
have no partners to which they can couple.

If the Higgs field $\phi_i$ acquires a non-zero vacuum expectation
value  $\phi^0_i$ by dynamical symmetry breaking,
then the fermion fields $\psi_{(j,p_i)\,\ell}$ and
$\widetilde\psi_{(j,p_i+2)\,\ell}$ acquire a mass matrix. For example, 
if one takes $k_0=k_1=\cdots =k_m=n$ and $\phi_i^0=\frac{1}{g\,R}\,v_i
~\Idd_n$ as in (\ref{phii0}), then the mass matrix for each
$\ell=0,1,\dots,2j$ is
\beq
\mbf M_f=\frac1R\,
\begin{pmatrix} \lambda_{j,p_i} & \frac{1}2\,v_i\\
\frac{1}2\,\overline v_i & \lambda_{j,p_i+2} \\
\end{pmatrix} \ ,
\eeq
with eigenvalues
\beq
\mu_\pm = \frac 1{2R}\,\left(\,\lambda_{j,p_i} +
  \lambda_{j,p_i+2}\pm\sqrt{\bigl(\lambda_{j,p_i} -
    \lambda_{j,p_i+2}\bigr)^2  + | v_i|^2}~\right) \ .
\label{mupmeigen}\eeq
These masses are proportional to $\frac1R$. In general, it seems
plausible that $\mu_-$ could be very small, or even zero, for specific
symmetry breaking patterns, but we have not found an example where
this happens. In this example all fermion fields transform in the
fundamental representation of the unbroken gauge group $\SU(n)$ after
spontaneous symmetry breaking.

Another interesting possibility arises when the metric on $M$ is of
lorentzian signature. Then all adjoint spinors $\psi^\dagger$ should
be replaced by $\overline\psi=R\, \psi^\dag \gamma^0$, where the
radius factor is necessary to maintain canonical dimensions in our
conventions since the gamma-matrix $\gamma^0$ has mass dimension one,
and similarly for $\widetilde\psi\,^\dag$. With the chiral
decompositions $\psi=\psi^+ \oplus\,\psi^-$ and
$\widetilde\psi=\widetilde\psi^{\,+}\oplus
\widetilde\psi^{\,-}$ on $M$ satisfying
\beq
\gamma\psi^\pm \=
\pm\,\psi^\pm \qquad \mbox{and} \qquad
\gamma\widetilde\psi^{\,\pm} \=
\pm\,\widetilde\psi^{\,\pm} \ ,
\label{gammapsis}\eeq
we are free to choose $\psi^-=
\widetilde\psi\,^+=0$ for the positive, negative and zero eigenvalues
of $\Dirac_{\C P^1}$. This makes the associated spinors
$\Psi^+_{(p_i)}$, $\Psi^-_{(p_i)}$, and $\Psi$
all Weyl fermions with positive chirality in $d+2$ dimensions.
The direct mass terms involving $\lambda_{j,p_i}$ in 
(\ref{nonnegative}) now all vanish leaving only Yukawa interactions
in $d$ dimensions.

\bigskip

\section{Dynamical symmetry breaking from the fundamental
  representation\label{fundrep}}

\noindent
We will now work through some explicit, illustrative examples of the
dimensionally reduced field theories of \S\ref{Eqgauge}. We will
obtain the complete physical particle content and compute all fermion
masses induced from the dynamical symmetry breaking. In this section
we will look at the gauge symmetry decomposition
(\ref{SUkgaugebroken}) in the case of restriction from the lowest
non-trivial $\su$ representation, the spin~$\frac12$
representation. This is the example $m=1$, which corresponds to the
fundamental representation of $\SU(2)$. A special instance of this
class of examples will involve an electroweak symmetry breaking
pattern.

\subsection{Higgs mechanism}

The gauge symmetry reduction is given by $\SU(k)\rightarrow
\SU(k_0)\times \SU(k_1)\times \uo$ with $k_0+k_1=k$. The $\uo$ factor
sits in the fundamental representation of $\SU(k)$ as the generator 
\beq
Y=\sqrt{\frac{2}{k}}\,\left(\begin{matrix}
    \sqrt{\frac{k_1}{k_0}}~{\mathbf 1}_{k_0} & {\mbf 0}_{k_0\times
      k_1} \\ 
{\mbf 0}_{k_1\times k_0} & -\sqrt{\frac{k_0}{k_1}}~{\mathbf
  1}_{k_1} \end{matrix}\right) \ ,
\eeq
where $\tr_{k\times k}(Y)=0$ and the normalisation is such that
$\tr_{k\times k}(Y^2)=2$. Here ${\mbf 0}_{k_0\times k_1}$ is the
$k_0\times k_1$ zero matrix.

The $\uo$ charge of the scalar field $\phi:=\phi_1$ follows from
(\ref{der}) with $i=0$ and the top left block of $Y$ acting on
$\phi$ from the left, as the $\uo$ part of $A^0$, while the bottom
right block of $Y$ acts on $\phi$ from the right, as the $\uo$ part
of $A^1$. This gives the $\uo$ charge of $\phi$ as
\beq
\sqrt{\frac{2}{k}}\,\left(\, \sqrt{\frac{k_1}{k_0}} +
  \sqrt{\frac{k_0}{k_1}}~\right)=\sqrt{\frac{2k}{k_0\,k_1}} \ .
\eeq
The bicovariant derivative (\ref{der}) can be written as
\beq 
D\phi = \diff\phi + \frac{\im g}{2}\,A_L^a\, \lambda_a \,\phi
-\frac{\im g}{2}\, A_R^{\tilde a} \,
\phi\, \tilde \lambda_{\tilde a}  +\frac{\im g}{2}\,
\sqrt{\frac{2k}{k_0\,k_1}}~B\,\phi \ ,
\eeq
where $\lambda_a$, $a=1,\ldots, k_0^2-1$ and $\tilde\lambda_{\tilde
  a}$, $\tilde a=1,\ldots, k_1^2-1$ are the Gell-Mann matrices for
$\SU(k_0)$ and $\SU(k_1)$ respectively, with $A_L^a$ and $A_R^{\tilde
  a}$ the corresponding left and right acting gauge fields, and $B$ is
the $\uo$ gauge field.

Without loss of generality we shall assume $k_0\ge k_1$. There is only
one Higgs multiplet $\phi$, which is a $k_0\times k_1$ complex matrix
field transforming under $\SU(k_0)$ from the left and $\SU(k_1)$ from
the right. The Higgs potential (\ref{Higgspot}) becomes 
\bea 
V(\phi) &=& \frac{g^2}2~\tr_{k_0\times k_0}\left( \frac{1}{4 g^2
    \,R^2} -\phi\,\phi^\dagger\right)^2 + \frac{g^2}2~\tr_{k_1\times
  k_1}\left( -\frac{1}{4 g^2 \,R^2} + \phi^\dagger\,\phi\right)^2
\nonumber\\[4pt] 
&=& \frac{k_0-k_1}{32 g^2 \,R^4}+g^2~\tr_{k_1\times
  k_1}\left(\frac{1}{4g^2\,R^2} -\phi^\dagger\, \phi\right)^2 \ .
\label{Vphi}\eea
We expect the gauge symmetry to be broken dynamically.

Using the $\SU(k_0)\times \SU(k_1)\times \uo$ gauge symmetry, a
generic $k_0\times k_1$ complex matrix $\phi$ can be brought into the
form
\beq 
\phi ~\longrightarrow~ U^{(0)} \,\phi \,U^{(1)}=
\frac 1 {gR}\left(\begin{matrix}  0 & 0 & \cdots & 0 \\
 & & \vdots & \\ 0 & 0 & \cdots & 0 \\ 
 v_1 & 0 & \cdots & 0 \\ 0 & v_2 & \cdots & 0 \\
 0 & 0 & \cdots & v_{k_1}\\ \end{matrix}\right) \ ,
\label{phiuntransf}\eeq
where $U^{(0)}$ is a $k_0\times k_0$ unitary matrix, $U^{(1)}$ is a
$k_1\times k_1$ unitary matrix, and $v_1,\ldots,v_{k_1}$ are
non-negative numbers. Putting the form (\ref{phiuntransf}) into the
potential (\ref{Vphi}), we find that the absolute minimum of $V(\phi)$
requires $v_1= \cdots = v_{k_1}=\frac{1}{2}$. Thus the vacuum
expectation value of $\phi$ is a bi-unitary transformation of the
matrix
\beq 
\phi^0 =\frac{1}{2g\,R}\,
\left(\begin{matrix} {\mbf 0}_{(k_0-k_1)\times k_1} \\
    \mathbf{1}_{k_1} \end{matrix}\right) \ .
\label{phi0}\eeq

The expectation value (\ref{phi0}) remains invariant under residual
$\SU(k_0 -k_1) \times \SU(k_1)_{\rm diag} \times \uo'$
transformations, where $\SU(k_1)_{\rm diag}$ is the diagonal subgroup
of the left and right acting groups $\SU(k_1)_L \times \SU(k_1)_R$
with $\SU(k_1)_L$ acting on the bottom $k_1$ rows of $\phi^0$, and
$\uo'$ is implemented as acting from the left on the top $k_0-k_1$
rows of $\phi^0$ leaving the bottom $k_1$ rows unchanged. This gives
the symmetry breaking pattern 
\beq
\SU(k_0)\times \SU(k_1)\times \uo~\longrightarrow~
\SU(k_0-k_1)\times \SU(k_1)_{\rm diag}\times \uo' \ .
\label{m1gensymbreak}\eeq
For the case $k_1=1$ the $\SU(k_1)$ factors are omitted.

A total of $2k_0\,k_1 - k_1^2$ gauge bosons acquire masses,
proportional to $\frac1R$, eating up $2k_0\,k_1 - k_1^2$ degrees of
freedom from the $k_0\times k_1$ complex matrix $\phi$ and leaving
$k_1^2$ physical Higgs fields. The latter fields can be arranged into
a $k_1\times k_1$ hermitean matrix $h=h^\dagger$ which sits in $\phi$
as 
\beq 
\phi=\left(\begin{matrix} {\mbf 0}_{(k_0-k_1)\times k_1} \\ 
\frac{1}{2g\,R}~\mathbf{1}_{k_1}+h\end{matrix}\right) \ , 
\label{vevplush}\eeq 
with $h$ an $\SU(k_0-k_1)$ singlet, transforming as a $k_1\times k_1$
hermitean matrix under the adjoint action of $\SU(k_1)_{\rm diag}$,
and carrying zero $\uo'$ charge. Expanding the potential (\ref{Vphi})
in powers of $h$ and examining the quadratic term reveals that the
Higgs bosons have mass $\mu_h=\frac{1}{R}$.

The precise masses of the gauge bosons can be determined by squaring 
(\ref{der}) for $i=0$, with $\phi_1=\phi^0$, and focusing on the part
quadratic in the gauge fields.  The mass matrix $\mbf {M}$ is a
symmetric matrix of dimension $(k_0^2 + k_1^2 -1)\times (k_0^2 + k_1^2
-1)$ and of rank $2k_0\,k_1-k_1^2$ defined via the relation
\beq 
\mbox{$\frac{1}{2}$}\, \mbf{A}^\top\, \mbf{M}^2\, {\mbf
  A}=g^2~\tr_{k_1\times k_1}
\left(A^0 \,\phi^0 - \phi^0\, A^1 \right)^\dagger\,
\left(A^0 \,\phi^0 - \phi^0\, A^1 \right) \ ,
\label{MassMatrix}\eeq
where $\mbf A$ is a column vector consisting of the $k_0^2 + k_1^2 -1$ 
gauge bosons in $\SU(k_0)\times \SU(k_1)\times \uo$. For example, if
$k_0=k_1=n$, then $\phi^0=\frac{1}{2g\,R}~{\mbf 1}_n$ and
\bea
\frac{1}{2}\, \mbf{A}^\top\, \mbf{M}^2 \,{\mbf A}&=&
\frac{g^2}{4}~\tr\left(A_L^a\, \lambda_a \,\phi^0 - A_R^a \,
\phi^0\, \lambda_a  + \frac{2}{\sqrt{n}}\,B\,\phi^0\right)^\dagger\,
\left(A_L^b\, \lambda_b\, \phi^0 - A_R^b \,
\phi^0\, \lambda_b\,  + \frac{2}{\sqrt{n}}
\,B\,\phi^0\right)\nonumber\\[4pt]
&=&\frac{1}{8R^2}\,\big(A_L \ \ A_R \ \ B\big)\,\left(\begin{matrix}
{\mbf 1}_{n^2-1} & -{\mbf 1}_{n^2-1} & 0 \\
-{\mbf 1}_{n^2-1} & {\mbf 1}_{n^2-1} & 0 \\
 0 & 0 & 2
\end{matrix}\right)\,\left(\begin{matrix} A_L \\ A_R \\
  B \end{matrix}\right) \ ,
\eea
where we have used the normalisation
$\Tr(\lambda_a\,\lambda_b)=2\delta_{ab}$. Diagonalising the mass
matrix we find $n^2-1$ massless gauge bosons
$A^a=\frac{1}{\sqrt{2}}\,(A^a_L + A^a_R)$, and $n^2-1$ massive vector
bosons $W^a=\frac{1}{\sqrt{2}}\,(A^a_L - A^a_R)$ which, together with
the $\uo$ gauge boson $B$, all have the same mass
$\mu^2_W=\mu^2_B=\frac{1}{2R^2}$.

An example which illustrates $\uo$ mixing is the case $k_1=1$. We
start with the simplest instance $k=3$, so that the gauge symmetry
reduction is $\SU(3)\rightarrow\SU(2)\times \uo$. The Higgs field
$\phi$ is \emph{a priori} a column vector with two complex
components. Let $\lambda_{\hat a}$, $\hat a=1,\ldots,8$ be the
Gell-Mann matrices generating $\SU(3)$, normalised so that
$\Tr(\lambda_{\hat a}\,\lambda_{\hat b})=2\delta_{\hat a\hat b}$. Then
the $\SU(2)$ generators can be chosen to be the Pauli spin matrices
$\sigma_a$, $a=1,2,3$ with 
\beq
\lambda_a=\left( \begin{matrix} \sigma_a & 0 \\ 0 & 0 \\
\end{matrix}\right) \ ,
\eeq
while the $\uo$ generator is 
\beq
\lambda_8=\sqrt{\frac{1}{3}}\,\left( \begin{matrix} {\mbf 1}_{2} & 0
    \\ 0 & -2 \\ \end{matrix}\right) \ .
\eeq
We thus set
\beq 
A^0 \=\frac{\im}{2}\,W^a\,\sigma_a + 
\frac{\im}{2\,\sqrt{3}}\, B ~{\mbf 1}_2 \qquad \mbox{and} \qquad
A^1\=-\im\,\sqrt{\frac{1}{3}} \,B \ ,
\eeq
where $W^a$ are the $\SU(2)$ gauge bosons and $B$ is the $\uo$ gauge
boson. The gauge coupling to $\phi$ now reads 
\beq
D\phi = \diff\phi + \frac{\im g}{2}\,\left( W^a\,\sigma_a + \sqrt{3}~
  B~ {\mbf 1}_2\right)\,\phi \ .
\eeq

We can use the $\SU(2)\times \uo$ gauge symmetry to rotate the
vacuum expectation value of $\phi$ to
\beq 
\phi^0=\frac{1}{2g\,R}\,
\left(\begin{matrix}  0 \\ 1\end{matrix}\right) \ ,
\eeq
and the quadratic form (\ref{MassMatrix}) becomes
\beq 
\frac{1}{2}\,
\big({\mbf W}^\top \ \ B\big)\,{\mbf
  M}^2\, \left(\begin{matrix} {\mbf W} \\ B\end{matrix}\right)
=\frac{1}{16 R^2}\,\left( W^a \,W^b\,\delta_{ab} +3 B^2
  -2\,\sqrt{3}~W^3\, B \right)
\eeq
with ${\mbf W}$ a three-component column vector. From this equation
we can read off the $4\times 4$ mass matrix
\beq 
{\mbf M}^2=\frac{1}{8 R^2}\,\left(\begin{matrix} 1 & 0 & 0 & 0 \\
 0 & 1 & 0 & 0 \\ 0 & 0 & 1 & -\sqrt{3} \\
0 & 0 & -\sqrt{3} & 3 
\end{matrix}\right) \ .
\eeq
The photon, i.e. the $\uo'$ gauge boson, is the massless combination
\beq
A=\mbox{$\frac{1}{2}$}\,\big(B+\sqrt{3}~ W^3 \big) \ ,
\eeq
while the $W$-bosons acquire a mass $\mu_{W^\pm}=\frac{1}{2\sqrt 2\,
  R}$ and the $Z$-boson a mass $\mu_Z=2\mu_{W^\pm}=\frac{1}{\sqrt 2
  \,R}$. Clearly this is not a realistic model for electroweak
interactions, since the Weinberg angle is too large at
$\sin^2\theta_{\rm  W}=\frac{3}{4}$. Nevertheless, it is an
instructive example for the symmetry hierarchy given by
$\SU(3)\rightarrow\SU(2)\times \uo\rightarrow \uo$ in this class of
models.

A similar analysis can be carried through for $k_0>2$ and $k_1=1$. 
In this case
$2k_0-1$ of the $\SU(k_0)$ gauge bosons acquire a mass while
$k_0^2-2k_0$ of them remain massless, and the residual gauge symmetry
is $\SU(k_0-1)$. There are $2k_0-1$ $W$-bosons, one of which mixes
with $B$ to form a massive $Z$-boson while leaving an orthogonal
linear combination massless. The mass matrix is of the form
\bea
{\mbf M}^2&=&\frac{1}{8R^2}\,
\left(\begin{matrix} 
{\mbf 0}_{k_0\,(k_0-2)\times k_0\,(k_0-2)} & 0 & \cdots & 0 \\
0 & {\mbf 1}_{2(k_0-1)} &0 & 0\\
\vdots & 0 & \frac{2(k_0-1)}{k_0} & -\frac{2\,\sqrt{k_0^2-1}}{k_0} \\
0 & 0 & -\frac{2\,\sqrt{k_0^2-1}}{k_0} & \frac{2(k_0+1)}{k_0} \\
\end{matrix}\right) \ . 
\eea
The gauge boson masses are independent of $k_0$. The $W$-boson mass is
$\mu_W=\frac{1}{2\sqrt{2}\,R}$, while diagonalising the bottom right
$2\times 2$ matrix block again reveals a $Z$-boson mass
$\mu_Z=\frac{1}{\sqrt{2}\,R}$. The same quantitative features hold for
any value of $k_0$ in the symmetry reduction sequence
\beq
\SU(2k_0)~\longrightarrow~ \SU(k_0)\times
\SU(k_0) \times \uo~\longrightarrow~ \SU(k_0)_{\rm diag} \times \uo' \
,
\eeq
with the ground state of the Higgs sector given by (\ref{phii0}). In
all instances we get the mass hierarchy
\beq 
\mu_h \= \sqrt 2 \,\mu_Z \= 2\sqrt 2 \,\mu_W \=\frac{1}{R} \ .
\eeq

\subsection{Yukawa interactions}

The $d$-dimensional fermion fields in this model are 
\beq
\widetilde\psi_{(1)} \ , \qquad  \psi_{(j,1)\,\ell} \ , \qquad
\mbox{and} \qquad \widetilde\psi_{(j,1)\,\ell}
\label{j0spinors}
\eeq
in the fundamental representation of $\SU(k_0)$, and 
\beq
\psi_{(-1)} \ , \qquad  \psi_{(j,-1)\,\ell} \ , \qquad \mbox{and}
\qquad \widetilde\psi_{(j,-1)\,\ell}
\label{j1spinors}
\eeq
in the fundamental representation of $\SU(k_1)$. In both cases $j\ge
1$ and $\ell=0,1,\ldots,2j$. The Yukawa couplings are obtained from 
(\ref{EFred}) and (\ref{nonnegative}), with $m=1$ and $i=0,1$. After
the rescalings (\ref{PhiPsiRescalings}), they are given by
\beq
S_{\rm Y}^{(j,\ell)}=\frac g2\,\int_M\,\diff^dx~\sqrt{|G|}~\left[
\left(\psi_{(j,-1)\,\ell}\right)^\dagger\,  \phi^\dagger \,\gamma\, 
\widetilde\psi_{(j,1)\,\ell} +
\big(\widetilde\psi_{(j,1)\,\ell}\big)^\dagger\,  \phi\, \gamma\, 
\psi_{(j,-1)\,\ell}\right]
\eeq
from the Dirac eigenspinors on $\C P^1$, and
\beq
S_{\rm Y}^0=\frac g2\,\int_M\,\diff^dx~\sqrt{|G|}~\left[
\left(\psi_{(-1)}\right)^\dagger\,  \phi^\dagger\, \gamma\, 
\widetilde\psi_{(1)} +
\big(\widetilde\psi_{(1)}\big)^\dagger \, \phi\, \gamma\, 
\psi_{(-1)}\right]
\eeq
from the zero modes. 

When the Higgs field $\phi$ is of the form (\ref{vevplush}),
the gauge symmetry is broken as in (\ref{m1gensymbreak}), and the
Yukawa couplings for the spinors (\ref{j0spinors}) and
(\ref{j1spinors}) are given by 
\bea
S_{\rm Y}&=& \frac{1}{4R}\,\int_M\,
\diff^dx~\sqrt{|G|}~\left[\left(\psi_{(j,-1)\,\ell}
\right)^\dagger\,\gamma\,\widetilde\psi_{(j,1)\,\ell} +
\left(\psi_{(-1)}\right)^\dagger\, \gamma\, 
\widetilde\psi_{(1)}\right]\nonumber\\
&&+\,\frac g2\,\int_M\,\diff^dx~\sqrt{|G|}~\left[
\left(\psi_{(j,-1)\,\ell}\right)^\dagger \,h \,\gamma\, 
\widetilde\psi_{(j,1)\,\ell} +\left(\psi_{(-1)}\right)^\dagger\,
h\,\gamma\,  
\widetilde\psi_{(1)}\right] +\mbox{h.c.} 
\eea
All fermion fields transform in the fundamental representation of
$\SU(k_1)_{\rm diag}$ after spontaneous symmetry breaking. There are
no Yukawa couplings for any of the fermion fields 
$\psi_{(j,1)}$ or $\widetilde\psi_{(j,-1)}$, although they pick up 
direct mass terms proportional to $\frac1R$ from the eigenvalues
$\lambda_{j,\pm\,1}=\sqrt{j\,(j+1)}$. The mass matrix for the fermions
$\psi_{(j,-1)}$ and $\widetilde\psi_{(j,1)}$ has eigenvalues
\beq
\mu_\pm=\frac1R\,\left(\,\sqrt{j\,(j+1)}\pm\mbox{$\frac14$}\,\right) \
. 
\eeq
For the fuzzy sphere truncation at $j_{\rm max}=0$, all spinors
$\psi_{(j,\pm \,1)\,\ell}$ and $\widetilde\psi_{(j,\pm \,1)\,\ell}$
vanish and only the zero modes survive, leaving two flavours of Dirac
fermions in $d$ dimensions, $\psi_{(-1)}$ and $\widetilde\psi_{(1)}$,
with Yukawa couplings and masses $\frac1{4R}$ but no direct mass
term.

\bigskip

\section{Dynamical symmetry breaking from the adjoint
  representation\label{adrep}}

\noindent
Our final example corresponds to the equivariant gauge theory which
descends from the spin one, adjoint representation of $\SU(2)$, having
$m=2$. This example constitutes the simplest case in which the
symmetry breaking is determined by a \emph{chain} of Higgs
multiplets. As in the previous section, we shall compute the complete
physical excitation spectrum in both the bosonic and fermionic sectors
for the spontaneously broken symmetry phase of the field theory.

\subsection{Higgs mechanism}

Let us consider the case in which $k=3n$ with $k_0=k_1=k_2=n$. Then
the gauge symmetry reduction sequence is
\beq
\SU(3n)~\longrightarrow ~\SU(n)_1\times \SU(n)_2\times \SU(n)_3\times
\uo\times \uo' \ .
\eeq
There are two complex $n\times n$ matrices of Higgs fields $\phi_1$,
with $\SU(n)_1$ acting on the left and $\SU(n)_2$ acting on the right,
and $\phi_2$, with $\SU(n)_2$ acting on the left and $\SU(n)_3$ on the
right. The $\uo$ and $\uo'$ generators can be taken to be any two
linearly independent generators of $\SU(3n)$ which commute with all
three $\SU(n)$ factors. In particular, we can take
\beq 
Y\=\frac{1}{\sqrt n}\,\left(\begin{matrix}
{\mbf 1} & {\mbf0} & {\mbf0} \\
{\mbf0} & -{\mbf 1} & {\mbf0} \\
{\mbf0} & {\mbf0} & {\mbf0} 
\end{matrix}\right)
\qquad \mbox{and} \qquad
Y'\=\frac{1}{\sqrt n}\,\left(\begin{matrix}
{\mbf0} & {\mbf0} & {\mbf0} \\
{\mbf0} & {\mbf 1} & {\mbf0} \\
{\mbf0} & {\mbf0} & -{\mbf 1}
\end{matrix}\right) \ ,
\eeq
where each entry is an $n\times n$ matrix. With this normalisation the
scalar field $\phi_1$ has $Y$-charge $\frac{2}{\sqrt n}$ and
$Y'$-charge $-\frac{1}{\sqrt n}$, while $\phi_2$ has $Y$-charge
$-\frac{1}{\sqrt n}$ and $Y'$-charge $\frac{2}{\sqrt n}$.

The Higgs potential (\ref{Higgspot}) is
\beq
V(\phi_1,\phi_2)=\frac{n}{4g^2\,R^4} - \frac{1}{2R^2}~\tr \left(
\phi_1^\dagger\, \phi_1 +\phi_2^\dagger\, \phi_2\right)
+g^2~\tr \left( \big(\phi_1^\dagger\,\phi_1\big)^2
-\big(\phi_1^\dagger\,\phi_1\,\phi_2^\dagger\,\phi_2\big)
+\big(\phi_2^\dagger\,\phi_2\big)^2\right) \ . 
\label{V3n}\eeq
The general solution (\ref{phii0}) in this case, with
$\zeta_1=\zeta_2=1$, gives the vacuum configuration
\beq
\phi_1^0\=\phi_2^0\=\frac{1}{\sqrt{2}\,g\,R}~{\mbf 1}_n \ .
\eeq
This expectation value is invariant under a single copy of $\SU(n)$,
which is a linear combination of $\SU(n)_1$, $\SU(n)_2$ and $\SU(n)_3$.
There is no linear combination of $\uo$ and $\uo'$ that leaves it
invariant, so the gauge symmetry is reduced as
\beq 
\SU(3n)~\longrightarrow ~
\SU(n)_1\times \SU(n)_2\times \SU(n)_3\times \uo \times \uo'
~\longrightarrow~ \SU(n) \ .
\eeq
In this case $2n^2$ gauge bosons acquire a mass and, of the $4n^2$
degrees of freedom in $\phi_1$ and $\phi_2$, a total of $2n^2$ remain
as physical Higgs fields. Expanding around the ground state, the
latter fields can be represented by $n\times n$ hermitean matrices
$h_1$ and $h_2$ with 
\beq 
\phi_1 \= \frac{1}{\sqrt{2}\,g\,R}~{\mbf 1}_n +h_1 \qquad
\mbox{and} \qquad
\phi_2 \= \frac{1}{\sqrt{2}\,g\,R}~{\mbf 1}_n +h_2 \ .
\label{phi12h12}\eeq
Putting this form into (\ref{V3n}) and examining the terms quadratic
in $h_1$ and $h_2$, we find a $2\times 2$ mass matrix with Higgs
masses given by the eigenvalues
\beq
\mu_{h_>}\=\frac{\sqrt 3} R \qquad \hbox{and} \qquad
\mu_{h_<}\=\frac{1}{R} \ .
\eeq

The vector boson masses are derived by diagonalising the
$(3n^2-1)\times (3n^2-1)$ mass matrix arising from the identity
\bea
&& \frac{1}{2}\, \mbf{A}^\top\, \mbf{M}^2\, \mbf{A} \\ && \=
\mbox{$\frac{1}{8R^2}$}\,\Tr
\left[
  \left( \big(A_1^a  - A_2^a\big)\, \lambda_a 
          +\mbox{$\frac{2}{\sqrt n}$}\, B\,{\mbf 1}_n
          -\mbox{$\frac{1}{\sqrt n}$}\, B'\,{\mbf 1}_n\right)^\dagger
        \, \left( \big(A_1^b - A_2^b\big)\, \lambda_b 
    +\mbox{$\frac{2}{\sqrt n}$}\, B\,{\mbf 1}_n -\mbox{$\frac{1}{\sqrt
        n}$}\, B'\,{\mbf 1}_n \right) \right.
\nonumber\\ 
&& \qquad +\,\left.\left( \big(A_2^a  - A_3^a)\, \lambda_a 
  -\mbox{$\frac{1}{\sqrt n}$}\, B\,{\mbf 1}_n
  +\mbox{$\frac{2}{\sqrt n}$}\, B'\,{\mbf 1}_n \right)^\dagger \,
\left( \big(A_1^b  - A_2^b\big)\, \lambda_b 
-\mbox{$\frac{1}{\sqrt n}$}\,B\,{\mbf 1}_n +\mbox{$\frac{2}{\sqrt
    n}$}\,B'\,{\mbf 1}_n\right) \right] \ , \nonumber
\eea
where $B$ and $B'$ are the gauge fields associated with $\uo$ and
$\uo'$ respectively. This gives
\beq
\mbf{M}^2=\frac{1}{4R^2}\,\left(
\begin{matrix} 2\,{\mbf 1} & -2\,{\mbf 1} & {\mbf 0}  & 0 &
  0 
  \\ -2\,{\mbf 1} & 4\,{\mbf 1} & -2\,{\mbf 1}  & 0 & 0 \\ 
{\mbf0} & -2\,{\mbf 1} & 2\,{\mbf 1}  & 0 & 0 \\ 
0 & 0 & 0 & 5 & -4 \\
0 & 0 & 0 & -4 & 5 \\
\end{matrix}\right) \ ,
\eeq
where each bold-face matrix block is of dimension
$(n^2-1)\times(n^2-1)$. The massless $\SU(n)$ gauge bosons are 
\beq 
A^a=\mbox{$\frac{1}{\sqrt 3}$}\,\big(A^a_1 + A^a_2 + A^a_3\big) \ ,
\eeq
while the vector bosons $W^a=\frac{1}{\sqrt 2}\,(A^a_1 - A^a_3)$
acquire a mass  
\beq
\mu_W=\frac{1}{\sqrt{2}\,R}
\eeq 
and $V^a=\frac{1}{\sqrt 6}\,(A^a_1 - 2 A^a_2 +A^a_3)$
have mass
\beq
\mu_V=\sqrt{\frac{3}{2}}~\frac{1}{R} \ .
\eeq
The $\uo$ vector bosons mix as
\beq
Z\=\mbox{$\frac{1}{\sqrt 2}$}\,\big(B+B'\,\big) \qquad \mbox{and}
\qquad Z'\=\mbox{$\frac{1}{\sqrt 2}$}\,\big(B-B'\,\big) \ ,
\eeq
with masses
\beq
\mu_Z\=\frac{1}{2R}\qquad \mbox{and} \qquad
\mu_{Z'}\=\frac32\,\frac{1}{R} \ .
\label{muZs}\eeq

\subsection{Yukawa interactions}

The $d$-dimensional fermion fields in this model are
\beq
\widetilde\psi_{(2)} \ , \qquad \psi_{(j,2)\,\ell} \ , \qquad \mbox{and}
\qquad \widetilde\psi_{(j,2)\,\ell} \qquad \mbox{with} \quad j\ge
\mbox{$\frac 32$} 
\eeq
in the fundamental representation of $\SU(n)_1$, together with
\beq
\psi_{(j,0)\,\ell} \qquad \mbox{and} \qquad
\widetilde\psi_{(j,0)\,\ell} \qquad \mbox{with} \quad j\ge \mbox{$\frac
  12$}
\eeq
in the fundamental representation of $\SU(n)_2$, and
\beq
\psi_{(-2)} \ , \qquad \psi_{(j,-2)\,\ell} \ , \qquad \mbox{and} \qquad 
\widetilde\psi_{(j,-2)\,\ell} \qquad \mbox{with} \quad j\ge \mbox{$\frac
  32$} 
\eeq
in the fundamental representation of $\SU(n)_3$. In all cases
$\ell=0,1,\dots,2j$. The Yukawa couplings after dimensional reduction
are
\beq
S_{\rm Y}=\frac g2\,\int_M\,
\diff^dx~\sqrt{|G|}~\left[
  \bigl(\widetilde\psi_{(j,2)\,\ell}\bigr)^\dag \,\phi_1\,
  \psi_{(j,0)\,\ell} 
+  \bigl(\widetilde\psi_{(j,0)\,\ell}\bigr)^\dag\, \phi_2\,
\psi_{(j,-2)\,\ell} \right]+ \mbox{h.c.} \ ,
\label{Yukawam2}\eeq
as the spinor fields $\psi_{(-2)}$, $\widetilde\psi_{(2)}$,
$\psi_{(j,2)\,\ell}$ and $\widetilde\psi_{(j,-2)\,\ell}$ have no
Yukawa couplings. The former two fermions are massless, while the
latter two fermions have direct mass terms proportional to $\frac1R$
coming from the eigenvalues
$\lambda_{j,\pm\,2}=\sqrt{(j-\frac12)\,(j+\frac32)}$. All fermion
fields transform in the fundamental representation of $\SU(n)$ after
spontaneous symmetry breaking. After substituting the Higgs fields
(\ref{phi12h12}) into (\ref{Yukawam2}), the Yukawa masses can be read
off from the general formula (\ref{mupmeigen}) with $p_i=0$
and~$|v_i|^2=\frac12$. 

\bigskip

\section{Conclusions\label{concl}}

\noindent
In this paper we have explicitly worked out the $\su$-equivariant
dimensional reduction of pure massless Yang-Mills-Dirac theory over
the coset space $\C P^1$, including a systematic incorporation of
Dirac monopole backgrounds. The internal magnetic fluxes induce a
Higgs potential as well as Yukawa couplings between the reduced
fermion fields and the Higgs fields, with the standard form of
dynamical symmetry breaking. In particular, in certain instances the
zero modes of the Dirac operator on $\C P^1$ acquire Yukawa
interactions. In our formulation we are able to
naturally induce both massive and massless fermions, as well as a
chiral gauge theory. When inducing massive Dirac spinors associated to
higher spinor harmonics on $\C P^1$, it is more natural to use a fuzzy
sphere $\C P^1_F$ for  the internal space, as it provides an
$\su$-equivariant truncation of the infinite tower of modes and
can also be used to truncate to the finitely-many flavours of massless
symmetric spinor modes. We worked out several explicit examples of
spontaneous symmetry breaking, including classes containing the
standard electroweak symmetry breaking sequence as a special case and
also a class involving a chain of Higgs fields. In all cases we
explicitly worked out the complete physical particle spectrum in the
dimensionally reduced field theory after dynamical symmetry breaking.

There are a few technical points which we have brushed over in our
analysis. For example, we have not analysed the stability of the Higgs
vacua $\phi_i^0$ that led to dynamical symmetry breaking. Although the
spectrum of fluctuations around the solutions we have used certainly
do not contain any unstable modes, because these vacua minimize the
Higgs potential, one should check whether or not there are any flat
modes which may lead to a non-trivial vacuum moduli space. This
appears to be a rather non-trivial task even for the simplest Higgs
vacua we have found. We have also not addressed the problem of
renormalizability of the dimensionally reduced field theory. Since the
original higher-dimensional Yang-Mills-Dirac theory is generically
non-renormalizable, keeping all higher modes in the
lower-dimensional model generically leads to a non-renormalizable
field theory. It is not clear if the truncations we have used can help
to give better quantum behaviour. It would be interesting to analyse
further if any symmetries of the coupled chain field system
(e.g. supersymmetry) could lead to renormalizable quantum field
theories after dimensional reduction.

In this article we have only focused on the simplest possible
homogeneous space to elucidate as clearly as possible the effects of
topologically non-trivial gauge field configurations obtained by
gauging the holonomy group of the coset. In principle, one can
consider more complicated coset spaces $G/H$ with the hope of
obtaining more realistic physical theories resembling the standard
model. As regards the fermionic sector, a particularly crucial role is 
played by those cosets which admit a finite-dimensional matrix
approximation $(G/H)_F$, such as the fuzzy complex projective spaces
$\C P_F^N$ where an explicit universal Dirac operator is known and
whose spectrum has been studied in detail in~\cite{DHMC}. For $N=2$,
the $\SU(3)$-equivariant dimensional reduction of Yang-Mills theory
over $\C P^2$ has been carried out in detail in~\cite{LPS3}
incorporating both $\SU(2)$ instanton and $\uo$ monopole backgrounds
associated with the holonomy group ${\rm U}(3)$ of $\C P^2$. It would
be interesting to extend the techniques of this paper to these classes
of equivariant dimensional reduction schemes. In particular, one can
compare with results of~\cite{DN} where the use of (fuzzy) complex
projective planes has been suggested as a natural internal space for
Kaluza-Klein reduction, leading to the appropriate chiral fermionic
spectrum of the standard model.

It would also be interesting to use our techniques to study the
reductions of the ten-dimensional ${\cal N}=1$ supersymmetric ${\rm
  E}_8$ gauge theories over six-dimensional coset spaces considered
in~\cite{KZ1,Lopes1}, although many of these cosets have no known
fuzzy versions. Nevertheless, the $\SU(3)$ equivariant dimensional
reduction of Yang-Mills theory over the six-dimensional non-symmetric
space $\SU(3)\big/\uo^2$ is explicitly worked out in~\cite{LPS3}
including $\uo$ monopole backgrounds associated with the maximal torus
$\uo\times\uo$ of $\SU(3)$. An outline of a scheme that could allow
for a fuzzy version of this coset space was proposed
in~\cite{saemann}. It would be interesting to compare the resulting
four-dimensional field theories with those of~\cite{Lopes1},
particularly the supersymmetry properties which arise under 
equivariant dimensional reduction. Our reduction techniques could also
be applied in principle to the superstring theories on nearly K\"ahler
backgrounds considered in~\cite{Lopes2}. In this regard it would be
interesting to find a natural interpretation for the internal fluxes
within the context of these superstring models, along the lines of the
flux stabilization mechanisms on arrays of D-branes in Type~II
string theory suggested in~\cite{LPS1,LPS2,PS1,LPS3}.

\bigskip

\section*{Acknowledgments}

\noindent
B.P.D. wishes to thank the Dublin Institute of
Advanced Studies for financial support. The work of R.J.S. was
supported in part by the EU-RTN Network Grant MRTN-CT-2004-005104.

\end{document}